\documentclass[structabstract]{aa}
\usepackage{graphicx}
\usepackage{rotating}
\usepackage{txfonts}
\usepackage{natbib}
\usepackage{lscape}
\usepackage{booktabs}
\usepackage{dcolumn}
\usepackage{float}
\usepackage{dblfloatfix}
\usepackage{array}
\usepackage{footnote}
\usepackage{adjustbox}
\makesavenoteenv{tabular}
\makesavenoteenv{table}
\usepackage{color}
\usepackage{hyperref}
\hypersetup{
  colorlinks=true,        
  linkcolor=blue,         
  citecolor=cyan,         
}

\bibpunct{(}{)}{;}{a}{}{,} 

\begin{document}
\renewcommand{\topfraction}{1.99}
\renewcommand{\bottomfraction}{1.99}
\renewcommand{\floatpagefraction}{1.99}
\renewcommand{\dbltopfraction}{1.99}
\renewcommand{\dblfloatpagefraction}{1.99}
\setcounter{totalnumber}{10}
\renewcommand{\textfraction}{1.99}

\newcommand{\cf}{cf.,~}
\newcommand{\ie}{i.e.,~}
\newcommand{\eg}{e.g.,~}
\newcommand{\etal}{\textit{et al.},}
\newcommand{\lr}[1]{\textcolor{red}{LR: #1}}

\newcommand{\cmf}[1]{\textcolor{blue}{CMF: #1}}
\newcommand{\zy}[1]{\textcolor{green}{ZY: #1}}

\def\simless{\mathbin{\lower 3pt\hbox
{$\rlap{\raise 5pt\hbox{$\char'074$}}\mathchar"7218$}}}   
\def\simmore{\mathbin{\lower 3pt\hbox
{$\rlap{\raise 5pt\hbox{$\char'076$}}\mathchar"7218$}}}   

\title{Using evolutionary algorithms to model relativistic jets}
\subtitle{Application to NGC\,1052}
\author{C. M. Fromm\inst{1,2}, Z. Younsi\inst{3,1}, A. Bazcko\inst{2}, Y. Mizuno\inst{1}, O. Porth\inst{4,1}, 
M. Perucho\inst{5,6}, H.\,Olivares\inst{1}, A. Nathanail\inst{1}, E.\,Angelakis\inst{2}, E. Ros\inst{2},  J. A. Zensus\inst{2} 
\and L. Rezzolla\inst{1,7}}

\institute{Institut f\"ur Theoretische Physik, Goethe Universit\"at,
  Max-von-Laue-Str. 1, D-60438 Frankfurt, Germany\
\and Max-Planck-Institut f\"ur Radioastronomie, Auf dem H\"ugel 69,
D-53121 Bonn, Germany\
\and Mullard Space Science Laboratory, University College London,
Holmbury St.\,Mary, Dorking, Surrey RH5 6NT, UK \
\and Anton Pannekoek Institute for Astronomy, University of Amsterdam,
Science Park 904, 1098 XH Amsterdam, The Netherlands\
\and Departament d'Astronomia i Astrof\'\i sica, Universitat de
Val\`encia, Dr. Moliner 50, E-46100 Burjassot, Val\`encia, Spain\
\and Observatori Astron\`omic, Parc Cient\'{\i}fic, Universitat de
Val\`encia, C/ Catedr\`atic Jos\'e Beltr\'an 2, E-46980 Paterna,
Val\`encia, Spain\
\and Frankfurt Institute for Advanced Studies, Ruth-Moufang-Strasse 1,
60438 Frankfurt, Germany
\email{cfromm@th.physik.uni-frankfurt.de}}
\abstract
{High-resolution Very-Long-Baseline Interferometry (VLBI) observations of NGC\,1052 show a 
two sided jet with several regions of enhanced emission and a clear emission gap between 
the two jets.
This gap shrinks with increasing frequency and vanishes around $\nu\sim43$\,GHz. 
The observed structures are due to both the macroscopic fluid dynamics interacting with 
the surrounding ambient medium including an obscuring torus and the radiation microphysics.
In order to model the observations of NGC\,1052 via state-of-the art numerical simulations
both the macro- and micro-physics have to be taken into account.}
{In this paper we investigate the possible physical conditions in relativistic jets of NGC\,
1052 by directly modelling the observed emission and spectra via state-of-the-art
special-relativistic hydrodynamic (SRHD) simulations and radiative transfer calculations.}
{We performed SRHD simulations of over-pressured and pressure-matched jets using the
special-relativistic hydrodynamics code \texttt{Ratpenat}. To investigate the physical 
conditions in the relativistic jet we coupled our radiative transfer code to evolutionary
algorithms and performed simultaneous modelling of the observed jet structure and the
broadband radio spectrum.
During the calculation of the radiation we consider both thermal and non-thermal 
emission. In order to compare our model to VLBI observations we take into account the 
sparse sampling of the u-v plane, the array properties and the imaging algorithm.}
{We present for the first time an end-to-end pipeline for fitting numerical simulations to VLBI 
observations of relativistic jets taking into account the macrophysics including fluid 
dynamics and ambient medium configurations together with thermal/non-thermal emission and 
the properties of the observing array.
The detailed analysis of our simulations shows that 
the structure and properties of the observed relativistic jets in NGC\,1052 can be reconstructed
by a slightly over-pressured jet ($d_k\sim1.5$) embedded in a decreasing pressure ambient
medium}
   {} \keywords{galaxies: active, -- galaxies: jets, -- radio continuum:
     galaxies, -- radiation mechanisms: non-thermal, radiative transfer, hydrodynamics}

\titlerunning{EA model-fitting of rel. jets}
\authorrunning{C. M. Fromm et al.}
\maketitle

\section{Introduction}
The giant elliptical galaxy NGC\,1052 (z=0.005037) harbours
a low-luminosity active galactic nucleus (AGN) in its centre. The strong optical emission lines in its 
spectrum classifies NGC\,1052 as a LINER \citep[see, \eg][]{2000ApJ...532..883G}.The radio structure at arcsecond scales reveal a two-sided structure, but the central core dominates the 
emission with about 85\% of the flux \citep{1984ApJ...284..531W}.  The two radio lobes have indications 
of hot spots, with an east-west orientation covering about 3\,kpc (projected).  The central radio source has 
a relatively flat spectrum at cm wavelengths with typical flux densities of 1$-$2\,Jy.  VLBI images show a 
milliarcsecond jet and counterjet extending over
15\,mas at cm wavelengths \citep[see, \eg][]{2003PASA...20..134K, 2004A&A...426..481K}. The jets are propagating in an east-west direction and are oriented at $\sim60^\circ$
in the sky (measured from North through East).Multi-frequency VLBI studies of NGC\,1052 display an emission gap between the eastern and the western jet, whereas the extent of this emission gap is decreasing with frequency
and disappears for $\nu\geqslant43$\,GHz \citep{2004A&A...426..481K}.
These observations are interpreted to be caused by a torus-like structure, so that not only synchrotron 
self-absorption is present, but also free-free absorptionThe extent of the torus can be studied via high-resolution multi-frequency VLBI observations, providing estimates between $0.1\,\mathrm{pc}$ and $0.7\,\mathrm{pc}$ \citep{2003PASA...20..134K}.Remarkably, the source also displays as well water maser emission \citep[see, \eg][]{2003ApJS..146..249B}  observed at positions aligned with the radio jet \citep{1998ApJ...500L.129C}. 

The source has shown prominent emission in x-rays which provide an estimate for the column density of 
$10^{22}\,\mathrm{cm}^{-2}$ 
to $10^{24}\,\mathrm{cm}^{-2}$ \citep{2004A&A...420..467K,2004A&A...426..481K}.
\citet{2002evn..conf..167K,2004A&A...420..467K} find indications of extended x-ray emission in addition 
to the nucleus in a {\it Chandra} image, in agreement with radio emission and optical emission from 
the {\it Hubble Space Telescope} \citep{2000ApJ...532..323P}.  
The X-ray spectra reveal a strongly absorbed continuum \citep[see, \eg][]{2000A&A...364L..80G} and 
hints of a relativistically broadened Fe\,K$\alpha$ emission originating from a few gravitational radii of the 
central object \citep{2009ApJ...698..528B}.  However,
\citet{2013PhDT.......479B} shows that the data of NGC\,1052 can also be modelled consistently without 
a broad  Fe\,K$\alpha$ line.Using VLBI monitoring of the source at 15\,GHz, 22\,GHz and 43\,GHz the dynamics and kinematics
at parsec-scales can be studied. The typical apparent speeds observed in NGC\,1052 are sub-luminal, 
with ranges between 0.25\,c and 0.5\,c.
\citet{2013PhDT.......479B} reports a detailed analysis of the MOJAVE observed images between 1995 
an 2012 with average sky motions of 0.74$\pm$0.06\,mas\,yr$^{-1}$ which corresponds to projected 
speeds of 0.230$\pm$0.011\,c, consistent with the values reported by \citet{2003A&A...401..113V} 
(0.73$\pm$0.06\,mas\,yr$^{-1}$), with no significant
differences between the values measured in jet and counter jet.

Based on VLBA observations at 43\,GHz between 2004 and 2009 \citet{2018Baczkosub} derived mean 
speeds of $0.343\pm0.037$\,c  in the western jet and  0.561$\pm$0.034\,c in the eastern jet.
Using the obtained kinematics the viewing angle, $\vartheta$, can be computed.
\citet{2003A&A...401..113V} report values for the viewing angle of $\vartheta \geq 57^\circ$, whereas the 
values of \citet{2013PhDT.......479B} constrain it to $\vartheta \geq 78.8^\circ$ for a v=0.238\,c (see 
its Sect. 3.1.3).  From the speeds measured in the jet and counter-jet, the upper limit of $\vartheta \leq 
85^\circ$ is consistent with the lower limit of 78.8$^\circ$. Due to significant differences between the eastern 
and western jet in flux density ratio as well as speed, the viewing angle based on the 43\,GHz VLBA 
observations could only be derived to lie between $60 \leq \vartheta \leq 90$ which is consistent with the 
aforementioned upper limit.

The study of the 43\,GHz structure of the jets shows a significant difference between the eastern and
western jet, and so the question arises as to whether the jets in NGC\,1052 appear asymmetric due
to the presence of an obscuring torus or if the jets are intrinsically asymmetric \ie asymmetries in
the jet launching/formation process close to the central black hole.
Given a viewing angle of nearly 90$^\circ$, NGC\,1052 is a perfect laboratory to investigate the
influence of the surrounding medium including the obscuring torus, the radiation micro-physics and
the jet launching mechanism.\\

To investigate the interplay between the non-thermal emission produced in the jets and 
the thermal absorption provided by the obscuring torus we perform 2D-axisymmetric SRHD simulations
of jets in a decreasing-pressure ambient medium and compute their radiative signatures.
Depending on the parameters of the torus \ie torus density, temperature and dimensions, flux density
asymmetries in the jets can be produced.
In addition, spectral indices, $\alpha>2.5$
are obtained within the region which is covered by the obscuring torus.
The imprint of the torus can also be found in the broadband radio spectrum, either as a flat or as a
double-humped spectrum \citep{2018A&A...609A..80F}.\\ 

In this paper we combine our jet-torus model with evolutionary algorithms (EA) and 
address the question of which kind of jet and torus configuration is required to best model the 
observed radio images and broadband spectrum of NGC\,1052.\\

The organisation of the paper is as follows. In Sect.~\ref{obs} we introduce the radio galaxy
NGC\,1052 and present stacked radio images and the average broadband spectrum of the source.
The numerical setup for the jet and the torus are introduced in Sect.~\ref{SRHDtorus} 
and a summary on the emission simulation is provided in Sect.~\ref{emission}.
The mathematical and numerical methods used during the optimisation process are 
explained in Sect.~\ref{opt}. 
The obtained results and their corresponding discussion can be found in Sect.~\ref{results}
and in Sect.~\ref{discussion}. 
Throughout the paper we assume an ideal-fluid equation of state $p=\rho\epsilon(\hat{\gamma}-1)$,
where $p$ is the pressure, $\rho$ the rest-mass density, $\epsilon$ the specific internal energy,
and $\hat{\gamma}$ the adiabatic index \citep[see, e.g.,][]{Rezzolla_book:2013}. 

The flux density errors, $\sigma S_\nu$, on the synthetic images are computed using a frequency-
dependent
calibration error $\sigma_{\rm cal}\sim0.1-0.14$ and the off-source rms: $\sigma S_\nu= \sigma_{\rm cal.}S_{\nu} 
+\rm{rms}$, where
$S_\nu$ is the flux density. We define the spectral index, $\alpha$, computed between two frequencies, 
$\nu_1$ and $\nu_2$ as:\\ 
$\alpha_{\nu_1,\nu_2}=\log_{10}\left(S_{\nu_1}/S_{\nu_2}\right)/\log_{10}\left({\nu_1/\nu_2}\right)$.

At its distance of $17.72\pm1.26$\,Mpc, an angular separation of
1\,mas corresponds to a projected length of about 9.5$\times$10$^{-2}$\,pc, and a proper motion of 
1\,mas\,yr$^{-1}$ 
corresponds to an apparent speed of 0.31\,c.

\section{Observations of NGC\,1052}
\label{obs}
NGC\,1052 is a frequently observed source within the MOJAVE survey
\footnote{\url{https://www.physics.purdue.edu/MOJAVE/}} \citep{2009AJ....137.3718L} and the FGamma 
program\footnote{\url{https://www3.mpifr-bonn.mpg.de/div/vlbi/fgamma/fgamma.html}} 
\citep{2016A&A...596A..45F}.
These surveys provide an excellent data base for both high-resolution radio images and densely
sampled multi-frequency single dish observations. 
NGC\,1052 frequently ejects new components into the jets which are propagating outwards, and during
the passage along the jet their flux density fades away.
Modelling the variation on top of the underlying steady-state flow requires the injection of perturbation
and increases the computational costs.
This complication can be circumvented by producing stacked radio images which smear out the variation
on top of the underlying flow and provide an average image of the source.
Given the wealth of available information on NGC\,1052, we produced stacked radio images at
22\,GHz and 43\,GHz together with the mean broadband radio spectrum\footnote{We only considered 
frequencies where at least 20 measurements are available}.
In the following we provide a short description of the data used in this work.

NGC1052 was observed 29 times between 2004 and 2009 with the VLBA at 22GHz and 43GHz. Details 
on data calibration and the imaging process can be found in \citet{2018Baczkosub}. The typical VLBA 
beam sizes are about $(0.5\times 0.2)\,$mas at 43\,GHz and $(0.8\times 0.3)\,$mas at 22\,GHz, the total 
flux densities during this interval are between 0.5 and 2.0\,Jy at both frequencies.  As described in 
\citet{2018Baczkosub} several properties as for example dynamics and distribution of Gaussian model fit 
parameters, including brightness temperatures and component sizes, at 43GHz have been analysed. 
Based on these properties the jets appear to be asymmetric. To produce stacked images the individual 
maps at 22\,GHz and 43\,GHz had been aligned on an optically thin feature at around -2\,mas distance 
to the west of the 22\,GHz map peak. The maps were restored with a common beam for all observations 
and stacked images for each frequency were produced by performing a pixel-by-pixel average over all maps, 
similar to the procedures described in \citet{2017MNRAS.468.4992P}.
\begin{table}[h!]
\centering
\begin{minipage}[t]{\columnwidth}
\caption{Image parameters for the stacked VLBA images}
\label{obspara}
\begin{tabular}{@{}c c c c c c c@{}} 
\hline\hline
$\nu$ & RMS$^{a}$& S$_\mathrm{peak}$ &S$_\mathrm{total}$& $\Theta_\mathrm{maj}$ & 
$\Theta_\mathrm{min}$ & P.A.\\
GHz& [mJy/beam] & [Jy/beam] & [Jy] & [mas] & [mas] & [$\deg$]\\
\hline 
\hline
22   & 0.39 -- 1 & 0.31 & 0.96 & 1.25 & 0.42 & $-$13.5 \\
43   & 0.63 -- 1& 0.32 & 0.79 & 0.72 & 0.27 & $-$11.9 \\
\hline
\multicolumn{7}{l}{$^a$ RMS values are determined in a region of the final map}\\
\multicolumn{7}{l}{ without significant source flux}\\
\end{tabular}
\end{minipage}
\end{table} 
\begin{figure}[h!]
\resizebox{8.8cm}{!}{\includegraphics{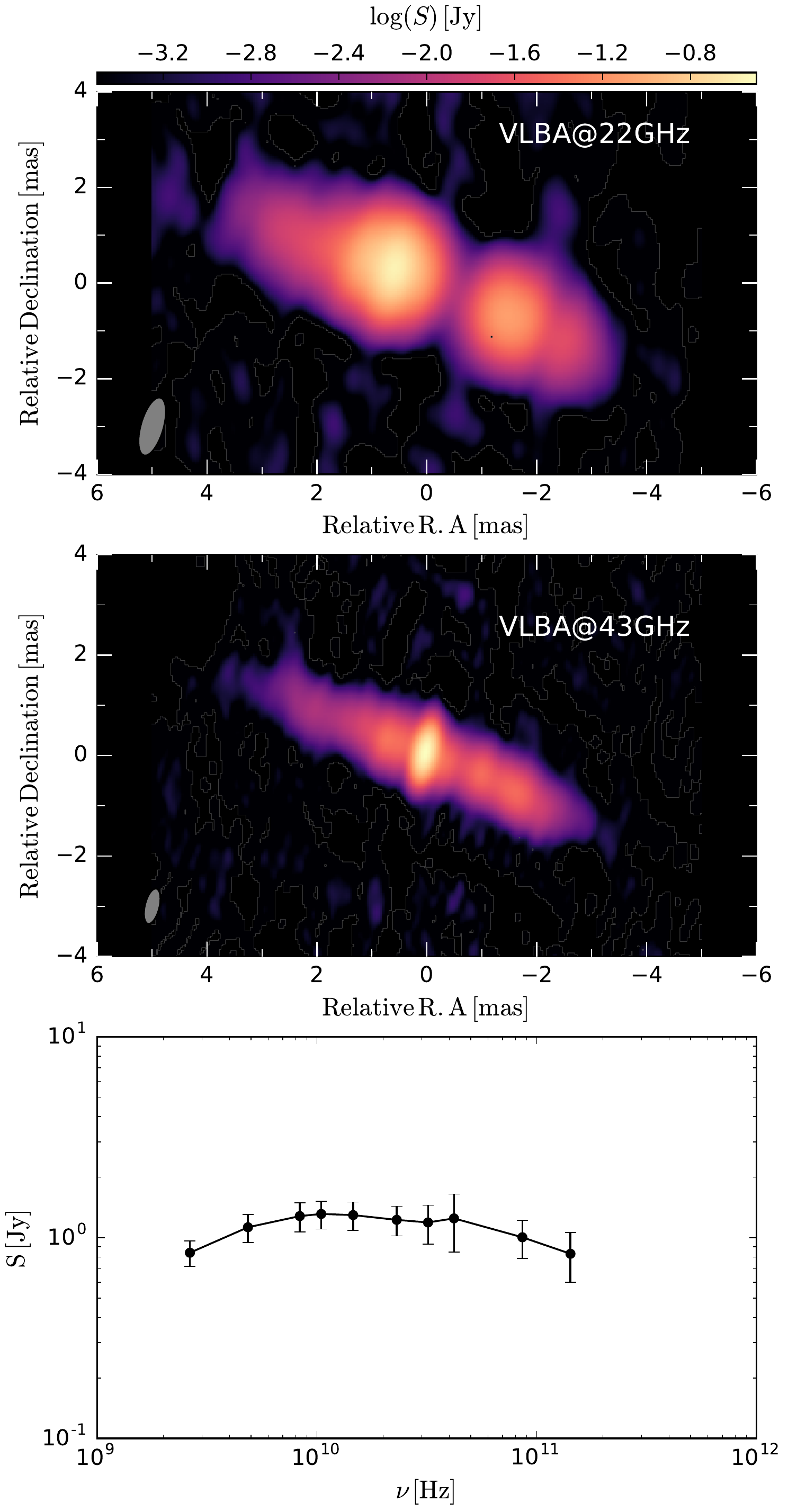}} 
\caption{Stacked VLBA images of NGC 1052 for 22\,GHz (top) and 43\,GHz (middle). The 
average radio spectrum for NGC\,1052 between 2.6\,GHz to 142\,GHz provided by FGamma 
is plotted in the bottom panel. For details on the image stacking see text.} 
\label{stackedimg} 
\end{figure}
%
\section{Numerical Simulations}
\label{modelsetup}
\subsection{SRHD and torus simulations}
\label{SRHDtorus}
\begin{figure*}[h]
\centering
\resizebox{14cm}{!}{\includegraphics{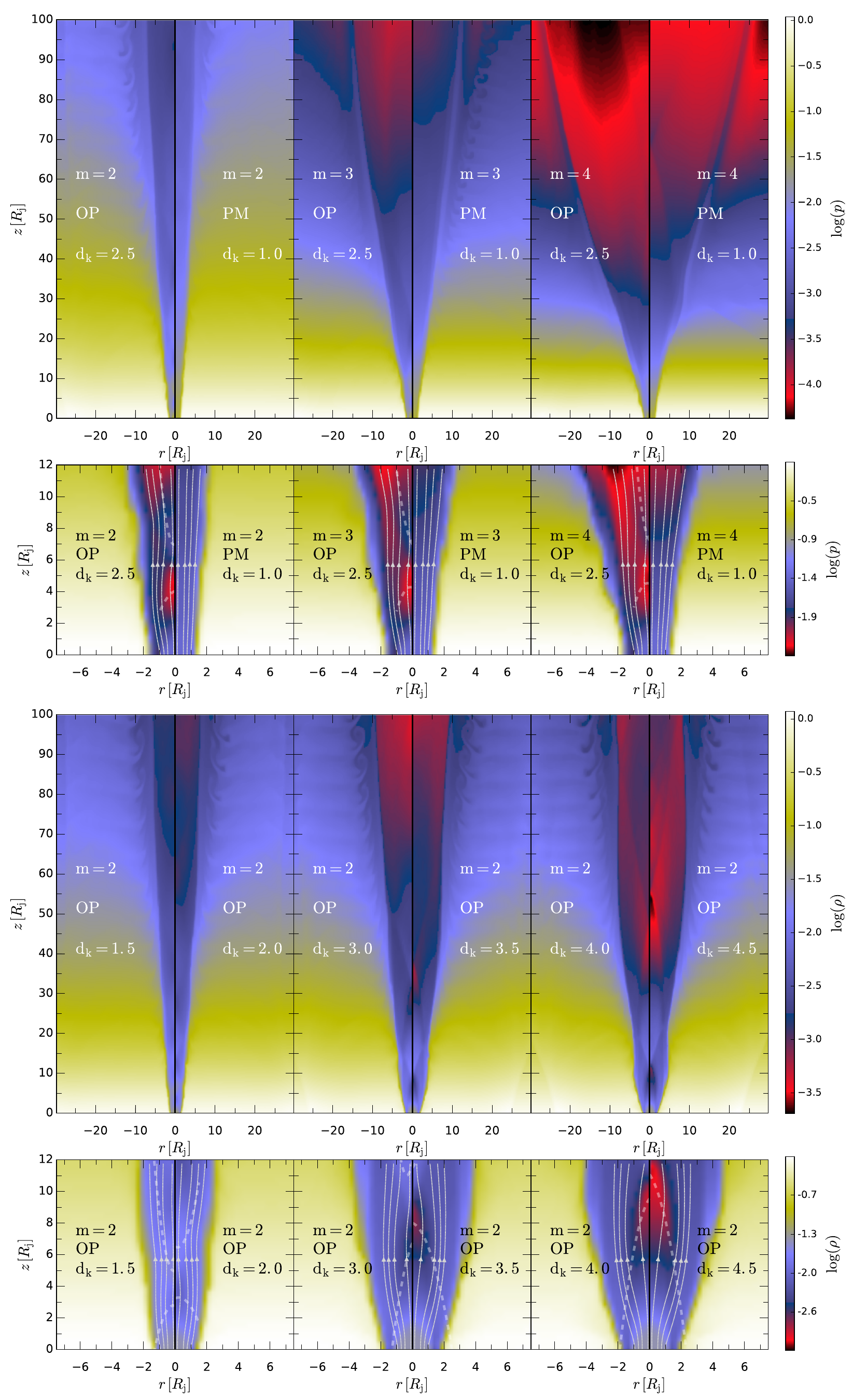}} 
\caption{Stationary results for the jet simulations.
The panels show the 2D distribution of the rest-mass density for different ambient medium
configurations as indicated by the exponent $m$ and $d_k$
($d_k=1$ corresponds to a PM jet and $d_k>1$ to an OP jet).
The upper row spans the entire simulation grid whereas the bottom row shows a magnified
view of the nozzle region ($-7 < z/R_{\rm j} < 7$).
The white lines in the bottom row show stream lines visualising the direction of the flow and
the bold dashed lines correspond to the inward travelling and reflected shock wave.}
\label{RHDall}
\end{figure*}

For our modelling of the jet in NGC\,1052 we use the SRHD simulations presented in 
\citet{2018A&A...609A..80F} and included additional values for the pressure mismatch,
namely $d_{k}=1.5,\,3.5,\,4.5\,\mathrm{and}\,5.0$.
For clarity we provide a short summary here.
The simulations are performed using 2D axisymmetric SRHD and the jets are injected in a
numerical grid which consists of 320 cells in the radial direction and 400 cells in the axial direction.
Using 4 cells per jet radius $(R_j)$ the numerical grid covers a range of $80\, R_j \times 100\,R_j$.
We inject jets with a velocity of $v_j=0.5\,\mathrm{c}$ in the z-direction at the left edge of the grid
$(z=0)$ and we assume a decreasing-pressure ambient medium.
The ambient medium follows a King-Profile which is characterised by the core-radius, $z_c$, and
the exponents $n$ and $m$ (see Eq. \ref{pamb}):
\begin{equation}
p_{\rm a}(z)=\frac{p_{\rm b}}{d_{\rm k}}\left[1+
\left(\frac{z}{z_{\rm c}}\right)^n\right]^{-\frac{m}{n}}\,.
\label{pamb}
\end{equation}
For the simulations presented in \citet{2018A&A...609A..80F} we applied $z_c=10\,R_{\rm j}$, 
$n=1.5$ and 
$m=2,3,4$. Depending on the ratio between the pressure in the jet and the ambient medium,
$d_k$, and the ambient pressure profile different jet morphologies are established (see 
Fig.~\ref{RHDall}).
If the jet is in pressure balance at the nozzle, nearly featureless conical jets are established,
so called pressure-matched (PM) jets.
On the other hand, a mismatch between the jet and ambient medium pressure at the nozzle leads to the
formation of a series of recollimation shocks and a pinching of the jet boundary is obtained.
These jets are referred to as over-pressured (OP) jets \citep[see, \eg][]{1997ApJ...482L..33G,Mimica:2009de,2006ApJ...640L.115A,2016A&A...588A.101F}\\
After roughly 5 longitudinal grid crossing times a steady state is recovered and we insert a
torus into our simulations.
This torus is modelled via several parameters which describe its geometry and the distribution of the
torus density and temperature \citep[see][for details]{2018A&A...609A..80F}. 
In Table \ref{paraem} we present an overview of the jet and torus parameters.
%
%
\begin{table}[h!]
\setlength{\tabcolsep}{4pt}
\caption{Parameters for the emission simulations. The $x$ indicates values which will be 
optimised during the modelling (see Sect. \ref{opt}).}
\label{paraem}
\centering
\begin{tabular}{@{}l l l@{}}
\hline\hline
Symbol  & Value &  Description\\
\hline
\multicolumn{3}{c}{\textit{\underline{scaling parameter}}}\\
$d_{\rm k}$			         &1,1.5, 2.5, 3.5,4.5,5								
		
	& \tiny pressure mismatch at nozzle\\
$R_{\rm j}$  				 &$3\times10^{16}$ cm							&  
\tiny jet radius	\\
$z_c$					& $10\,R_{\rm j}$								
	& \tiny core radius\\
$n$						 &1.5											
	& \tiny pressure gradient in ambient medium\\
$m$						&2, 3, 4										
	& \tiny pressure gradient in ambient medium\\
$z$						 &0.005										
	& \tiny redshift\\
$v_{\rm j}$				 &$0.5\,\mathrm{c}$								& 
\tiny jet velocity\\
$\hat{\gamma}$			 &13/9										& 
\tiny adiabatic index \\ 
$\rho_{\rm a} $				 &$x$ $\,\mathrm{g \, cm^{-3}}$	       		        & \tiny 
ambient medium density	\\
\hline \hline
\multicolumn{3}{c}{\textit{\underline{emission parameter}}}\\
$\epsilon_{\rm B}$			 &$x$											& 
\tiny equipartition ratio\\
$\epsilon_{\rm e}$ 		        &$x$											
	& \tiny thermal to non-thermal energy ratio\\
$\zeta_{\rm e}$ 		       &$x$											& 
\tiny thermal to non-thermal  number density ratio\\
$\epsilon_\gamma$		 &1000											& 
\tiny ratio between e$^-$ Lorentz factors\\
$s$						 &$x$											& 
\tiny spectral index\\
$\vartheta$				 &80$^\circ$ 									
	& \tiny viewing angle\\
\hline \hline
\multicolumn{3}{c}{\textit{\underline{torus parameter}}}\\
$L_{\rm AGN}$		&$1\times10^{43}\,\mathrm{erg \,s^{-1}}$			& \tiny bolometric 
luminosity\\
$R_{\rm out}$		&$x$ $\,\mathrm{cm}$					& \tiny torus outer radius\\
$\theta$					&$x$ $^\circ$ 									
	& \tiny torus thickness\\
$\rho\left(R_{\rm in}\right)$	 &$x$ $\,\mathrm{g \, cm^{-3}}$		& \tiny torus density at  
$R_{\rm min}$\\
$T_{\rm sub}$			 &$x$ K									& \tiny dust 
sublimation temperature\\
$k$, $l$					&$x$					 					&\tiny 
exponents for $\rho$ and $T$ distributions\\
\hline\hline
\end{tabular}
\end{table}
\subsection{Emission Calculations}
\label{emission}
In order to compare our SRHD models to the observations \ie VLBI images and single dish 
spectra, we need to compute the non-thermal and thermal emission.
For the computation of the emission we follow the recipe presented in \citet{2018A&A...609A..80F}.
For completeness we introduce the basic steps for the emission simulation below.
Since we are not evolving the non-thermal particles during the SRHD simulations we have to reconstruct 
their distribution from the SRHD variables \ie pressure, $p$ and density $\rho$.
We assume a power law distribution of relativistic electrons:
\begin{equation}
n\left(\gamma_{\rm e}\right)=n_0\left(\frac{\gamma_{\rm e}}{\gamma_{\rm e,\,min}}\right)^{-s}
\quad \mathrm{for\,\,\,} \quad
\gamma_{\rm e,\,min}\leq\gamma_{\rm e}\leq\gamma_{\rm e,\, max} \,,
\end{equation}
where $n_0$ is a normalisation coefficient, $\gamma_{\rm e}$ is the electron
Lorentz factor, $\gamma_{\rm e,\, min}$ and $\gamma_{\rm e,\, max}$ are the lower
and upper electron Lorentz factors and $s$ is the spectral slope.
In the next step we relate the number density of relativistic particles to the
number density of thermal particles in the jet via the scaling parameter $\zeta_{\rm e}$ as:
\begin{equation}
\int_{\gamma_{\rm e,\,min}}^{\gamma_{\rm e,\,max}}n\left(\gamma_{\rm e}\right) d \gamma=
\zeta_{\rm e}\frac{\rho}{m_{\rm p}} \,,
\label{nnumber}
\end{equation}
with $m_{\rm p}$ the mass of the proton. The energy of the non-thermal particles is related 
to 
the energy of the thermal particles via the parameter $\epsilon_{\rm e}$:
\begin{equation}
\int_{\gamma_{\rm e,\, min}}^{\gamma_{\rm e,\,max}}n\left(\gamma_{\rm e}\right)
\gamma_{\rm 
e}
m_{\rm e} c^2  d \gamma =
\epsilon_{\rm e}\frac{p}{\hat{\gamma}-1},
\label{nenergy}
\end{equation}
By performing the integrals in Eq. \ref{nnumber} and Eq. \ref{nenergy} an expression for the 
lower electron Lorentz factor, $\gamma_{\rm e,min}$ can be derived:

\begin{equation}
\gamma_{\rm e,\,min}=\left\{ \begin{array}{ll} 
\displaystyle
\frac{p}{\rho} \frac{m_{\rm p}}{m_{\rm e}
  c^2}\frac{(s-2)}{(s-1)(\hat{\gamma}-1)}\frac{\epsilon_{\rm e}}{\zeta_{\rm e}}
& \textrm{if }s>2 \,, \vspace*{3mm}\\
\displaystyle
\left[\frac{p}{\rho} \frac{m_{\rm p}}{m_{\rm e}
    c^2}\frac{(2-s)}{(s-1)(\hat{\gamma}-1)}\frac{\epsilon_{\rm e}}{\zeta_{\rm e}}
  \gamma_{\rm e,\,max}^{s-2}\right]^{1/(s-1)} & \textrm{if }1<s<2
\,, \vspace*{3mm}\\
\displaystyle
\frac{p}{\rho}\frac{\epsilon_{\rm e}}{\zeta_{\rm e}}
\frac{m_{\rm p}}{m_{\rm e} c^2(\hat{\gamma}-1)} \left[
  \ln\left(\frac{\gamma_{\rm e,\,max}}{\gamma_{\rm e,\,min}}\right)
  \right]^{-1} & \textrm{if } s=2 \,.
\label{gmineq1}
\end{array}\right. 
\end{equation}
The last step in the construction of the non-thermal particle distributions assumes that 
the 
upper electron Lorentz factor, $\gamma_{\rm e,max}$, is a constant fraction
$\epsilon_{\gamma}$ of the lower electron Lorentz factor
\begin{equation}
\gamma_{\rm e,\,max}=\epsilon_\gamma\,\gamma_{\rm e,\,min} \,.
\label{gmax}
\end{equation}
Given the expressions for the electron Lorentz factors (Eq. \ref{gmineq1} and \ref{gmax}) the 
normalisation coefficient of the particle distribution, $n_0$, can be written as:
\begin{equation}
n_0=\frac{\epsilon_{\rm e} p (s-2)}{\left(\hat{\gamma}-1
  \right)\gamma_{\rm  e,\,min}^2 m_e c^2}\left[ 1-\left(\frac{\gamma_{\rm e,\,
      max}}{\gamma_{\rm e,\,min}}\right)^{2-s}\right]^{-1} \,.
\label{nnorm}
\end{equation}
Given that information on the magnetic field cannot be obtained from our
hydrodynamical numerical simulations, we assume its magnitude is a
fraction $\epsilon_{\rm B}$ of the equipartition magnetic field
\begin{equation}
B=\sqrt{8\pi\epsilon_{\rm B}\frac{p}{\hat{\gamma}-1}}\,.
\label{Bcal}
\end{equation}
Finally, we compute the frequency-dependent total intensity, $I_\nu$, for each ray
by solving the radiative transport equation
\begin{equation}
\frac{ d I_\nu}{ d s}=\epsilon_{\nu,\mathrm{nt}} -
\left(\alpha_{\nu,\mathrm{nt}} + 
\alpha_{\nu,\mathrm{th}}\right)I_\nu\,,
\end{equation}
where $\epsilon_{\rm nt}$ and $\alpha_{\rm nt}$ are the emission and absorption coefficients 
for non-thermal emission and $\alpha_{\rm th}$ is the absorption coefficient for thermal 
emission \citep[for details on the computation of the emission and absorption coefficients 
see][]{2018A&A...609A..80F} .
\subsection{Modification of the emission calculations}
\subsubsection{Ray-tracing through an adaptive grid}
\label{adapgrid}
\begin{figure*}[h!]
\centering
\resizebox{17cm}{!}{\includegraphics{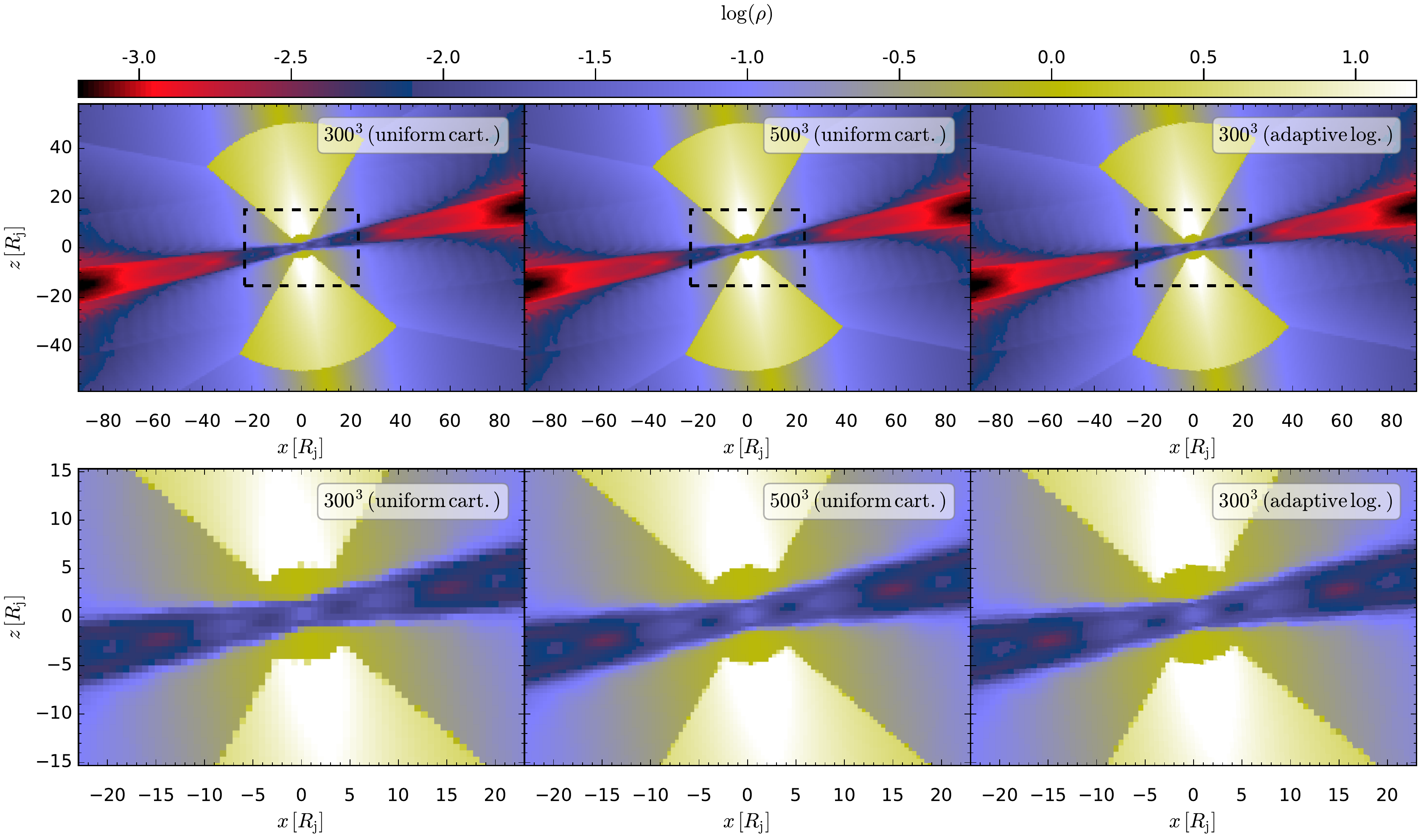}} 
\caption{Rest-mass density distribution for different grid resolutions (left: uniform cartesian 
$300^3$, center: uniform cartesian $500^3$ and right: adaptive logarithmic $300^3$). The 
top row shows entire simulation box and the bottom row a zoom into the central region 
indicated by the dashed boxes in the top row. The adaptive grid is created using the 
parameters presented in Table \ref{gridpara}.}
\label{linlogimages}
\end{figure*}

In our modelling we will compare VLBI images and single dish spectra at various frequencies. 
Assuming that the emission at higher frequencies is generated mainly near the apex of the jet 
(due to the higher density and magnetic field within this region as compared to the more 
extended outer regions), we have to ensure that we sufficiently cover this region within 
our numerical grid during the ray-tracing.
On the other hand, the low frequency emission zones are not restricted to a region close to
the apex of the jet. 
Therefore a minimum numerical resolution has to be provided to resolve the larger extent of the jet.
In order to fulfil our above stated criteria on the ray-grid we would need a high numerical resolution
which leads to computationally challenging simulations.
A way to overcome these computational limitations is to introduce adaptive non-linear grids.
The advantage of these grids, as compared to uniform Cartesian grids, is that we can increase
the numerical resolution at the apex of the jet and at the same time provide enough resolution
at the larger scales.
In this way we can keep the computational effort to a minimum (see Fig.~\ref{linlogimages}). 

In order to focus numerical resolution on the jet we first have to align the initial Cartesian 
coordinate system, indicated by the subscript ``$\mathrm{cart}$", with the jet direction.
The aligned coordinate system labeled with the subscript ``$\mathrm{align}$" is obtained from
the Cartesian system via two rotations using the angles $\vartheta$ (viewing angle) and
$\varphi$ (orientation of the jet in plane of the sky).
After the rotation into the aligned system the coordinates are modified according to:
\begin{eqnarray}
x_{\rm align}&=&\mathrm{sign}\left(x_{\rm cart} \right) \left ( \Delta x_{\rm align,\,min} +\frac{|
x_{\rm cart}|}{x_{\rm scale}} \right )^{\eta_x} \,, \\
y_{\rm align}&=&\mathrm{sign}\left(y_{\rm cart} \right) \left ( \Delta y_{\rm align,\,min} +\frac{|
y_{\rm cart}|}{y_{\rm scale}} \right )^{\eta_y} \,, \\
z_{\rm align}&=&\mathrm{sign}\left(z_{\rm cart} \right) \left ( \Delta z_{\rm align,\,min} +\frac{|
y_{\rm cart}|}{z_{\rm scale}} \right )^{\eta_z} \,,
\end{eqnarray}
where $\Delta (x,y,z)_{\rm align,\,min}$ is the smallest cell spacing in the different directions, 
$(x,y,z)_{\rm scale}$ sets the extent of the linear scale \ie number of cells covered with the 
highest numerical resolution and $\eta_{x,y,z}$ scales the exponential growth of the grid.
Once this ray grid is established we interpolate the SRHD parameters to each cell using a
Delaunay triangulation.
The parameters for the numerical grid used in this work are $\Delta (x,y,z)_{\rm align,\,min}=10^{-3}$,  
$(x,y,z)_{\rm scale}=4$ and $\eta_{x,y,z}=1.15$.

%
%
%
%
%
A ray propagating through this non-linear grid will encounter cells with different cell sizes and 
we therefore have to take special care in the computation of the optical depth along ray paths. 
Consequently, we interpolate the SRHD values required for the computation of the
emission along the ray and include a bisection method for the calculation of the optical depth 
with an accuracy of $\Delta \tau \simeq10^{-6}$.
This method guarantees that we are tracing the optical depth cut-off with high precision and
leads to converged spectra and images (see Appendix for a detailed study on the spectra and
image convergence). 
\subsection{Synthetic imaging}
\begin{figure}[h!]
\resizebox{\hsize}{!}{\includegraphics{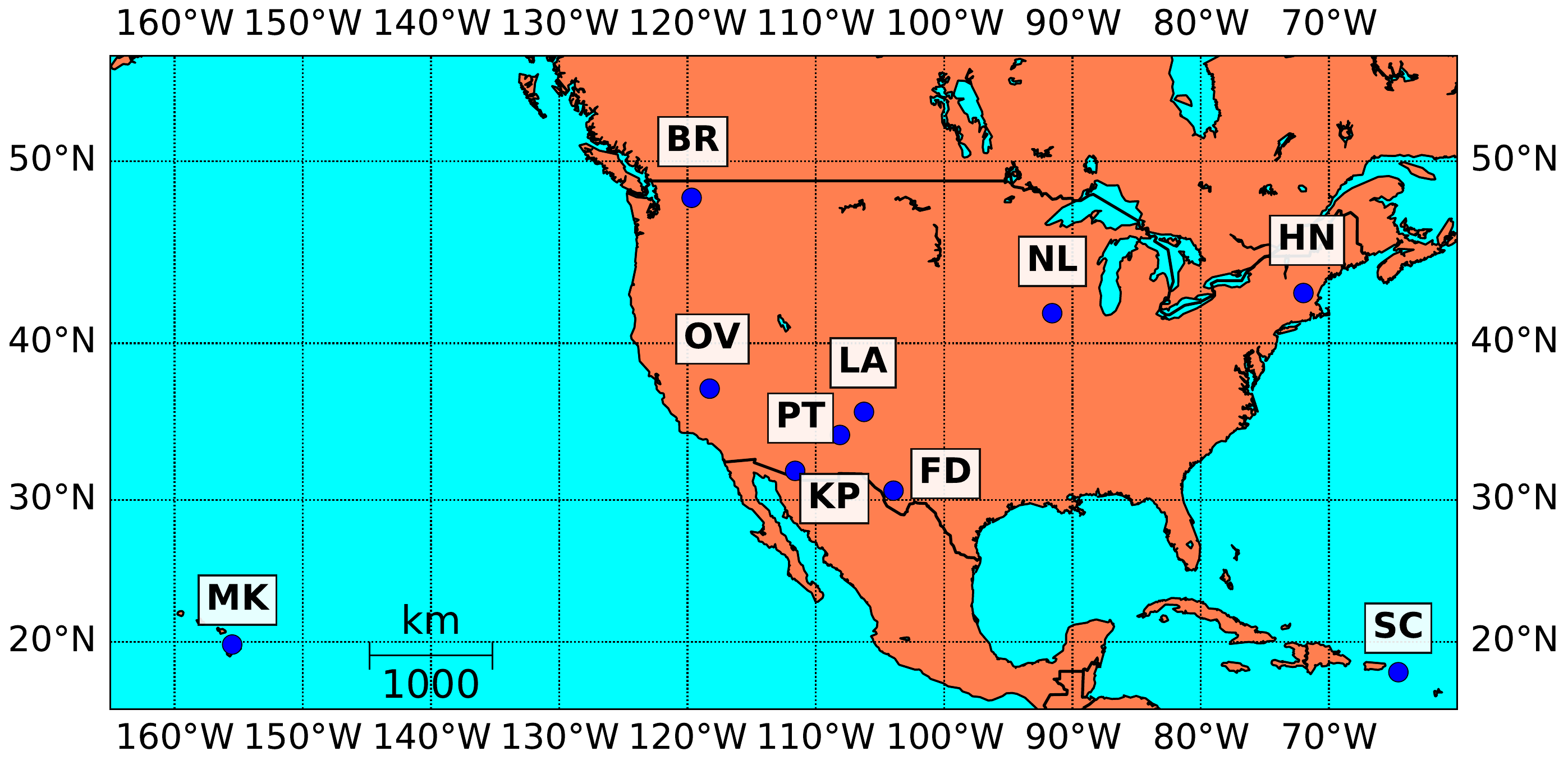}} 
\caption{Location of the VLBA antennas across the USA}
\label{VLBAsites} 
\end{figure}

A typical VLBI experiment consists of series of on-source scans on the main target and off-
source scans for calibration and focussing on a calibrator source.
In addition to the reduced on-source time due to calibration of the experiment, the limited number
of telescopes participating in the observations lead to a sparse sampling of the source brightness 
distribution in Fourier space (hereafter u-v plane).
Both effects reduce the number of data points in Fourier space (hereafter termed visibilities).
During the standard emission calculations these effects are not considered and the obtained
images are typically blurred with the observing beam to mimic real observations.
However, if we want to compare our emission simulations directly to VLBI observations we have
to take the above mentioned effects into account.
The calculation of the synthetic observations can be divided into four main steps:
\begin{enumerate}
\item Radiative transfer calculation.
\item Setup of the observing array and the observation schedule (duration of scans, integration time).
\item Fourier transformation of the computed intensity and sampling with the projected 
baselines of the observing array.
\item Reconstruction of the image.
\end{enumerate}
The solution to the radiative transfer problem is presented in Sect.~\ref{emission} and below 
we provide some details regarding the remaining steps listed above.\\
\paragraph{Array setup and observation schedule:}
Throughout this work we use the Very Long Baseline Array (VLBA) as the observing array.
The VLBA consists of 10 identical 25\,metre radio telescopes scattered across the USA
(see Fig.~\ref{VLBAsites}).
The antennas are equipped with several receivers allowing multi-frequency observations
between 1.6\,GHz to 43\,GHz (eight of the ten telescopes can also observe at 86\,GHz).
The sensitivity of a telescope is usually given in the system equivalent flux density (SEFD)
which is computed from the system temperature, $T_{\rm sys} $, and the effective 
area of the telescope, $A_{\rm eff}$\footnote{the effective area is the product of the 
geometrical telescope area and the aperture efficiency}.
In Table \ref{antennapara} we list the SEFD used for our synthetic imaging.
The SEFDs of a baseline (telescope 1 and telescope 2) together with the bandwidth,
$\Delta \nu$, and the integration time, $t_{\rm int}$, determine the thermal noise in the
synthetic images according to:
\begin{equation}
\sigma=\frac{1}{0.88}\sqrt{\frac{\mathrm{SEFD_1}\times\mathrm{SEFD_2}}{2\Delta \nu 
t_{\rm int}}} \,.
\end{equation} 
We use a bandwidth of 256\,MHz and an integration time of 10\,s for all frequencies listed in 
Table \ref{antennapara}.
In order to include the calibration gaps in the synthetic radio observations we assume
10 minutes on-source and 50 minutes off-source per observing hour.
Together with an integration time of 10\,s this transforms into $\sim60$ data points per 
scan per baseline (the exact number depends on rise and set time of the source). 
Additionally we include the effects of 10\% gain calibration errors following \citet{2016ApJ...829...11C}.
\begin{table}[t!]
\caption{Used SEFD for the VLBA}  
\label{antennapara}
\centering
\begin{tabular}{l|c|c|c|c|c|c|}
\hline\hline
frequency [GHz] & 1.6 & 4.8 & 8 & 15 & 22 & 43\\ 
\hline
SEFD [Jy] & 289 & 278 & 327 & 543 & 640 & 1181\\
\hline
\multicolumn{7}{l}{data taken from NRAO}
\end{tabular} 
\end{table} 
\paragraph{Fourier transformation and sampling:}
The computed intensity distribution is Fourier transformed and sampled with the projected 
baselines for the VLBA array.
Given that each telescope has certain elevation constraints together with the coordinates
of the source in the sky (RA and DEC) and the date of the observation, the source is not
always visible for the entire array.
We apply a lower elevation cutoff of $10^\circ$ an upper limit of $85^\circ$.
The synthetic data is generated using the 
\texttt{EHTim library}\footnote{\url{https://github.com/achael/eht-imaging}} which we 
modified to suit our needs. 
\paragraph{Image reconstruction:}
The simulated visibilities are imported into \texttt{DIFMAP} \citet{Shepherd:1997p2298} and 
the image is reconstructed using the \texttt{CLEAN} algorithm \citet{1974A&AS...15..417H} 
combined with the \texttt{MODELFIT} algorithm. The final image is stored in \texttt{FITS} 
format for further analysis \eg comparison between synthetic and observed images via 
normalised cross correlation coefficients and other image metrics.
\begin{figure}[h!]
\resizebox{\hsize}{!}{\includegraphics{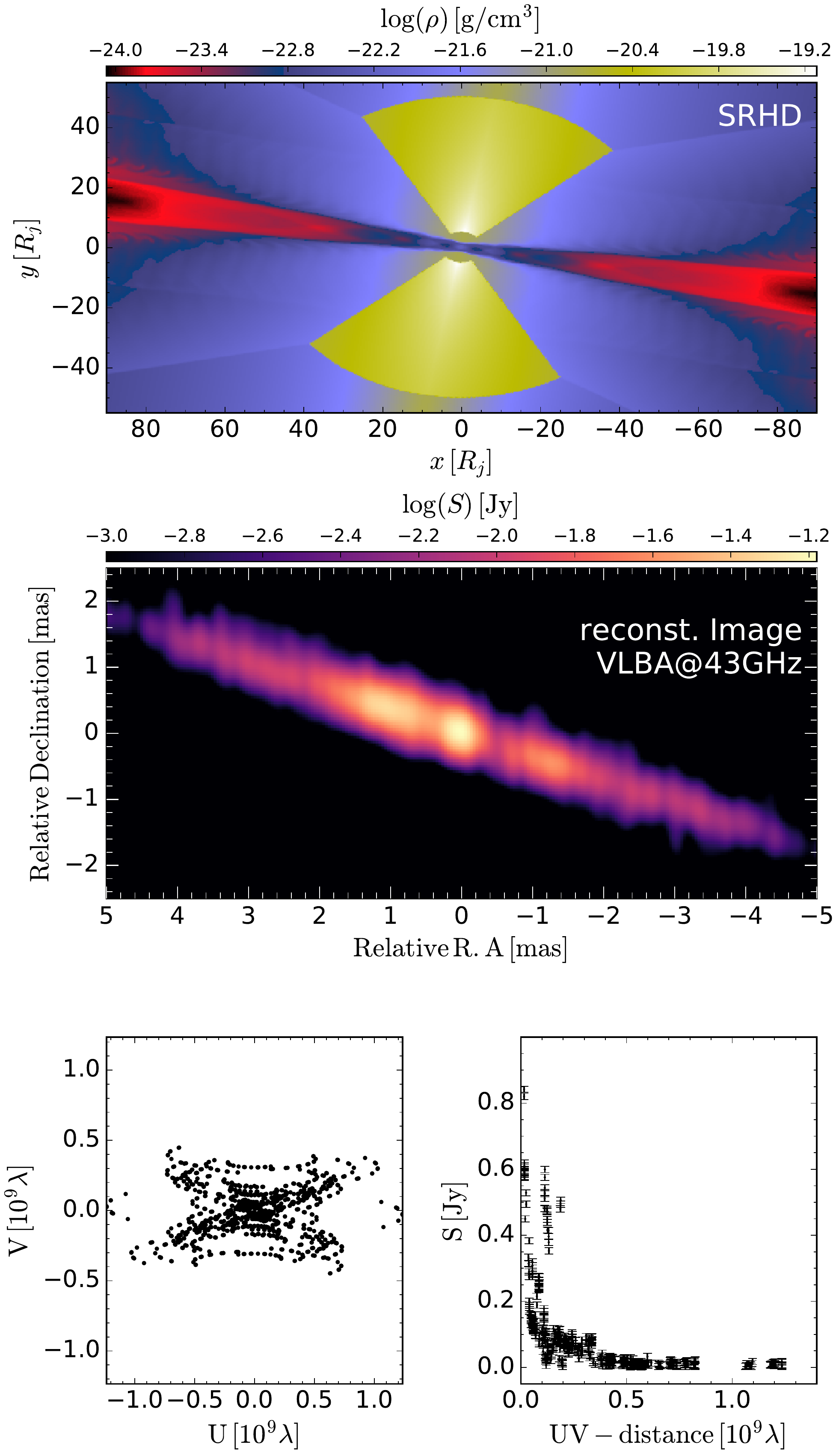}} 
\caption{Example from the synthetic imaging routine.
The top panel shows the logarithm of the rest-mass density in $g/cm^3$,
the middle panel displays the reconstructed radio image at 43\,GHz as seen by the VLBA
and the bottom panels present the u-v plane (left) and the visibility amplitude (right).
See text for details on the jet model used.}
\label{synexample} 
\end{figure}
%
In Fig.~\ref{synexample} we show an example from the synthetic imaging routine.
The underlying SRHD simulation and emission model parameters are summarised in Table \ref{paraemsyn}. 
The location of the source in the sky is 2h41m4.799s in R.A. and 
-8d15\arcmin20.752\arcsec in DEC.
We observe the source from 2017-04-04 0:00 UT to 2017-04-04 24:00 UT using the recipe
and array configuration mentioned above.
\begin{table}[h!]
\setlength{\tabcolsep}{4pt}
\caption{Parameters used for the calculation of the synthetic image presented in
Fig.~\ref{synexample}.}
\label{paraemsyn}
\centering
\begin{tabular}{@{}l l l@{}}
\hline\hline
\small  Symbol & Value &  Description\\
\hline
\multicolumn{3}{c}{\textit{\underline{scaling parameter}}}\\
$d_{\rm k}$			         &2.5										
	& \tiny pressure mismatch at nozzle\\
$R_{\rm j}$  				 &$3\times10^{16}$ cm	&\tiny jet radius	\\
$z_c$					& $10\,R_{\rm j}$								
	& \tiny core radius\\
$n$						 &1.5											
	& \tiny pressure gradient in ambient medium\\
$m$						&2									
	& \tiny pressure gradient in ambient medium\\
$z$						 &0.005										
	& \tiny redshift\\
$v_{\rm j}$				 &$0.5\,\mathrm{c}$								& 
\tiny jet velocity\\
$\hat{\gamma}$			 &13/9										& 
\tiny adiabatic index \\ 
$\rho_{\rm a} $				 &$1.67\times10^{-21}\,\mathrm{g/cm^{3}}$	       		        
& \tiny 
ambient medium density	\\
\hline \hline
\multicolumn{3}{c}{\textit{\underline{emission parameter}}}\\
$\epsilon_{\rm B}$			 &0.2											& 
\tiny equipartition ratio\\
$\epsilon_{\rm e}$ 		        &0.4											
	& \tiny thermal to non-thermal energy ratio\\
$\zeta_{\rm e}$ 		       &1.0											& 
\tiny thermal to non-thermal  number density ratio\\
$\epsilon_\gamma$		 &1000											& 
\tiny ratio between e$^-$ Lorentz factors\\
$s$						 &2.2											& 
\tiny spectral index\\
$\vartheta$				 &80$^\circ$ 									
	& \tiny viewing angle\\
\hline \hline
\multicolumn{3}{c}{\textit{\underline{torus parameter}}}\\
$L_{\rm AGN}$		&$1\times10^{43}\,\mathrm{erg \,s^{-1}}$			& \tiny bolometric 
luminosity\\
$R_{\rm out}$		&$1.5\times10^{18}\,\mathrm{cm}$					& \tiny torus 
outer radius\\
$\theta$					&40$^\circ$ 									
& \tiny torus thickness\\
$\rho\left(R_{\rm in}\right)$	 &$8.3\times10^{-20}\,\mathrm{g/cm^{3}}$		& \tiny 
torus density at  
$R_{\rm min}$\\
$T_{\rm sub}$			 &1400 K									& \tiny dust 
sublimation temperature\\
$k$, $l$					&1,2					 					&\tiny 
exponents for $\rho$ and $T$ distributions\\
\hline\hline
\end{tabular}
\end{table}
\section{Constrained non-linear optimisation} 
\label{opt}
For applying our model to the observational data we need to find the best values
for the different parameters listed in Table \ref{paraem}.
This task can be considered as a constrained non-linear optimisation problem:
\begin{eqnarray}
\begin{array}{ll@{}ll}
\mathrm{minimize}  & \displaystyle  f(\vec{x}) &\\
\mathrm{subject\, to}& \displaystyle g_{j}(\vec{x})\leq 0, &   &j=1 ,..., n\\
                 &  \displaystyle        x_{L,i}\leq x_{i}\leq x_{R,i},  & &i=1 ,..., m,
\end{array}
\end{eqnarray}
where $\vec{x}$ is an $m$ dimensional vector including the model parameters, \eg
$\vec{x}=\left[j_{\rm mod} ,\rho_a,\epsilon_b,\epsilon_e,\zeta_e,s,R_{\rm out},\theta,
\rho(R_{\rm in}), T_{\rm sub},k,l\right]^{T}$ and
$f(\vec{x})$ is the objective function (minimisation function), $g_j(\vec{x})$ are the 
constraints and
$x_{L,i}$ and $x_{R,i}$ are lower and upper boundaries for the model parameters.\\
The minimisation function and the constraints can be constructed in such a way that key 
properties of the data are used to guide the optimisation process which will speed up the 
convergence of the algorithm.
In the case of NGC\,1052 we use the number of local flux density maxima along the jet
axis and the existence/non-existence of an emission gap between the western jet
and the eastern jet.\\
For the minimisation function we use the least squares computed from the flux density
along the jet axes, $\chi^2_{\rm{ridge,i}}$ and the least squares computed from the
observed radio spectrum and simulated one, $\chi^2_{\rm spec}$:
\begin{equation}
f\left(\vec{x}\right)=-\left\{\left[\sum_{p=1}^q w_{i}\,\chi^2_{\rm{ridge,i}}(\vec{x})\right]
+w_{2q+1}\,\chi^2_{\rm spec}(\vec{x})\right\}^{-2} \,,
\label{f2}
\end{equation}
with $q$ denoting the number of images (frequencies) included in the data set with
weighting factors $w_{i}$.
The above mentioned constraints can be expressed as:
\begin{eqnarray}
\label{ccconst}
g_{i=1,\cdots, q}(\vec{x})&=&0.8 -{\rm cc_i(\vec{x})}\\
g_{i=q+1,\cdots,2q+1}(\vec{x})&=&n(\vec{x})_{\rm peak,i-q}^{\rm obs}-n(\vec{x})_{\rm peak,i-
q}^{\rm sim} \label{npconst}\\
g_{i=2q+1,\cdots,3q+1}(\vec{x})&=& \Delta r(\vec{x})_{\rm gap,i-2q}^{\rm obs}-\Delta 
r(\vec{x})_{\rm gap,i-2q}^{\rm sim} \label{gapconst}
\end{eqnarray}
The first constraints (Eq.~\ref{ccconst}) set the required minimum cross correlation coefficient 
between the observed and the synthetic radio images, \ie rejecting or accepting solutions 
based on the structural agreement between the images.
Equation~\ref{npconst} increases the agreement between the number of the flux density peaks,
$n(\vec{x})_{\rm peak,p}^{\rm obs\,sim}$, in the radio maps and Eq.~\ref{gapconst} is enforcing
a consistence between the images regarding a possible emission gap and its extent,
$\Delta r(\vec{x})_{\rm gap,p}^{\rm obs,\,sim}$.
During the optimisation process we allow for all constraints (eqns.~\ref{ccconst}--\ref{gapconst}) 
a tolerance of 0.01 \ie a cross correlation coefficient of ${\rm cc_{p}(\vec{x})=0.79}$ is accepted
as a solution.\\The numerical handling of the constraints depends on the implementation of the optimisation 
algorithm. A common approach includes the addition of penalty functions to the minimisation function, $f(\vec{x}$). 
More details on the constraint implementation can be found in \citep{Deb:2002,Jansen:2011}.
\subsection{Optimisation Algorithms}
The optimisation problem can be solved by two kind of algorithms: gradient-based and 
gradient-free algorithms.
Given the high dimensionality of our problem, together with the high computational effort
(ray-tracing and synthetic imaging) a gradient-based algorithm will most likely get stuck in a
local minimum and requires a lot of computational resources for mapping out the gradient with 
sufficient resolution.
Therefore, we apply a gradient-free search algorithm.
Among these classes of algorithm we select a genetic algorithm (GA) and a particle swarm
optimisation (PSO).
In the next Section we provide a short introduction to GA and PSO algorithms and refer 
to \citet{CI_book:2007} for further details.
\subsection{Genetic algorithm (GA)}
GAs are motivated by Darwin's theory of the survival of the fittest.
The main steps of a GA are ranking, crossover (mating), and mutation of the individuals
for several generations.
An individual can be seen as set of parameters, in our case $\vec{x}=\left[j_{\rm mod} ,\rho_a,
\epsilon_b,\epsilon_e,\zeta_e,s,R_{\rm out},\theta, \rho(R_{\rm in}), T_{\rm sub},k,l\right]^{T}$, 
also referred to as a chromosome, and each entry \eg $\theta$ is labelled as a gene.
During the initial step, $N$ random chromosomes are produced and their fitness is computed
(in our case the fitness function is given by Eq.~\ref{f2}).
Based on their fitness the chromosomes are ranked and selected for crossover.
During the crossover new chromosomes are created from the parent and the fitness of the
offsprings is computed.
If the fitness of the offspring is improved compared to the parent, the worst parent with respect
to the minimisation function can be replaced (details of crossover and replacing methods depend
on the implementation of the GA).\\
In addition to the crossover process, mutation is used to enforce diversity into the population.
Mutation is applied to the offspring of the crossover process at a certain rate, typically $\ll 1$,
which guarantees that good offsprings are not overwritten. 
During the mutation, one or more genes within a chromosome are selected and replaced, where 
again details on the mutation depend on the implementation of the GA.
To summarise, the main parameters of a GA are the number of chromosomes, the number of
generations (iterations), the fitness function, the rate of crossover and the rate of mutation.
In this work use the Non Sorting Genetic Algorithm II (\texttt{NSGA2}) \citep{Deb:2002}. 
\subsection{Particle swarm optimisation (PSO)}
A PSO mimics the behaviour of animal swarms, e.g., birds searching for food.
A swarm particle can be described by its position vector $\vec{x_i}$  and 
velocity $\vec{v_i}$ in the parameter space, and the food can be related to the minimisation 
function, here Eq.~\ref{f2}.
The optimisation is driven by the velocity $\vec{v_i}$, which reflects both the individual experience
of the particle, commonly referred to cognitive component, and the social experience,
\ie exchange of information between neighbouring particles.
Thus, the update on the position of each particle can be written as:
\begin{equation}
\vec{x_i}(t+1)=\vec{x_i}(t)+\vec{v_i}(t+1) \,.
\end{equation}
Depending on the variant of the PSO, different recipes for the calculation of the velocity exist
\citep[see Chapter 16 in][for details]{CI_book:2007}.
In the so-called global best PSO, the velocity update of a particle, $\vec{v_i}(t+1)$, is computed
from the best position found by the entire swarm at the time $t$ and the best position visited so
far by the individual particle (i).
The weighting between the best position of the swarm and the best position of the individual
particles is set by the social and cognitive weights.
The PSO terminates after a maximum number of iterations or after a convergence of the solution,
\ie $\Delta f\left(\vec{x}\right) <\epsilon$, where $f$ is the minimisation function.
In this work we make use of the Augmented Lagrangian Particle Swarm Optimisation
(\texttt{ALPSO}) \citep{Jansen:2011}. 
\subsection{Markov Chain Monte Carlo (MCMC) simulation }
To further investigate the uncertainties of optimal solutions found by the GA or the PSO we 
perform MCMC simulations using \texttt{emcee} \citep{2013PASP..125..306F}.
As initial positions for the random walkers we employ a Gaussian distribution with a mean
equal to the solution found by the GA or PSO and 50\% standard deviation.
The large standard deviation allows the random walkers to spread out and sufficiently
sample the parameter space around the PSO/GA position.
This hybrid approach is advantageous in that we don't have to use large burn-in times during 
the MCMC simulations, significantly reducing the computational effort and speeding up the 
calculation. 
We typically apply 400 random walkers and perform $10^{3}$ iterations, which leads to a 
total number of $4\times10^{5}$ iterations, similar to the number of iterations used by the GA
and the PSO runs.
\subsection{Optimisation strategy}
In order to model the observations of NGC\,1052 we employ the following strategy:
\begin{itemize}
\item Use different SRHD models presented in Table \ref{paraem} and Fig.~\ref{RHDall}.
\item Start the optimisation using GA or PSO, typically between $10^4$ and $10^5$ iterations
\item Use the broadband radio spectrum to guide the optimisation.
\item Explore the parameter space around the best position provided by GA or PSO via 
MCMC simulations.
\end{itemize}

The computational costs for a single run depends on the number of frequency points included in the broadband radio spectrum
and on the required step size to achieve the optical depth accuray, $\Delta \tau$. Typically 100\,s are required to compute a radio spectrum
consisting of 10 frequencies. This leads to a total required computational time of 300 to 3000 cpus hours. Using MPI parallelisation the duration of the computation can be reduced to a few tens of hours.
We perform a parameter recovery test to explore the capabilities of our end-to-end pipeline 
by inserting the reference model into our code.
The parameters of the reference model are well recovered within $1\sigma$ and the scatter
of the solutions provides insights into the uncertainties of the method (see Appendix for details).
\section{Results}
\label{results}
In Table \ref{parabest} we present the results of the optimisation runs, for both OP ($d_k>1$)
and PM ($d_k=1$) jets.
Both optimisation runs used $~4\times10^4$ iterations and results of the MCMC can be found
in the Appendix.
\begin{table}[h!]
\setlength{\tabcolsep}{4pt}
\caption{Best fit parameters obtained from non-linear optimisation}
\label{parabest}
\centering
\begin{tabular}{@{}l l l@{}}
\hline\hline
Symbol  & OP jets &  PM jets\\
\hline
\multicolumn{3}{c}{\textit{\underline{scaling parameter}}}\\
$d_{\rm k}$		& 1.5 &1.0\\
$R_{\rm j}$  		&\multicolumn{2}{c}{$3\times10^{16}$ cm} \\
$z_c$			& \multicolumn{2}{c}{$10\,R_{\rm j}$}\\
$n$				&\multicolumn{2}{c}{1.5}\\
$m$				&\multicolumn{2}{c}{2}\\
$z$				&\multicolumn{2}{c}{0.005}\\
$v_{\rm j}$		&\multicolumn{2}{c}{$0.5\,\mathrm{c}$}\\
$\hat{\gamma}$	&\multicolumn{2}{c}{13/9}\\ 
$\rho_{\rm a} $		&$3.0\times10^{-21}\,\mathrm{g \, cm^{-3}}$	    &$2.1\times10^{-21}\,
\mathrm{g \, cm^{-3}}$ \\
\hline \hline
\multicolumn{3}{c}{\textit{\underline{emission parameter}}}\\
$\epsilon_{\rm B}$			 &0.39 	&0.20 \\
$\epsilon_{\rm e}$ 		        &0.27		&0.34 \\
$\zeta_{\rm e}$ 		       &0.35        &0.40 \\
$\epsilon_\gamma$		 &1000		& 1000 \\
$s$						 &3.8		& 3.9\\
$\vartheta$				 &80$^\circ$ 	& 80$^\circ$ \\
\hline \hline
\multicolumn{3}{c}{\textit{\underline{torus parameter}}}\\
$L_{\rm AGN}$		&\multicolumn{2}{c}{$1\times10^{43}\,\mathrm{erg \,s^{-1}}$}\\
$R_{\rm out}$		&$1.4\times10^{18}\,\mathrm{cm}$					& 
$1.5\times10^{18}\,\mathrm{cm}$\\
$\theta$					&73$^\circ$ 	& 54$^\circ$  \\
$\rho\left(R_{\rm in}\right)$	 &$1.1\times10^{-19}\,\mathrm{g \, cm^{-3}}$		& 
$2.6\times10^{-20}\,\mathrm{g \, cm^{-3}}$\\
$T_{\rm sub}$			 &1250 K									& 1160 K\\
$k$, $l$				 &2.4 (2.3), 2.5 (1.5)					 			         & 2.2 
(1.6), 3.8 (1.5) \\
\hline\hline
\\[-1em]
\multicolumn{3}{c}{\textit{\underline{$\chi^2$ results}}}\\
\\[-1em]
$\chi^2_{\rm 22\,GHz}$ & 4.72 & 3.88 \\ 
\\[-1em]
$\chi^2_{\rm 43\,GHz}$ & 7.30 & 9.90 \\
\\[-1em]
$\chi^2_{\rm SED}$     & 1.28 & 1.94 \\
\\[-1em]
\hline \hline
\\[-1em]
\multicolumn{3}{c}{\textit{\underline{image metrics}}}\\
\\[-1em]
$\mathrm{cc}_{\rm 22\,GHz}$ & 0.98 & 0.98 \\ 
\\[-1em]
$\mathrm{cc}_{\rm 43\,GHz}$ & 0.91 & 0.85 \\
\\[-1em]
$\mathrm{DSSIM}_{\rm 22\,GHz}$     & 0.22 & 0.21 \\
\\[-1em]
$\mathrm{DSSIM}_{\rm 43\,GHz}$  &0.10   & 0.11 \\
\\[-1em]
\hline \hline
\end{tabular}
\end{table}
\begin{figure*}[t]
\centering
\resizebox{18cm}{!}{\includegraphics{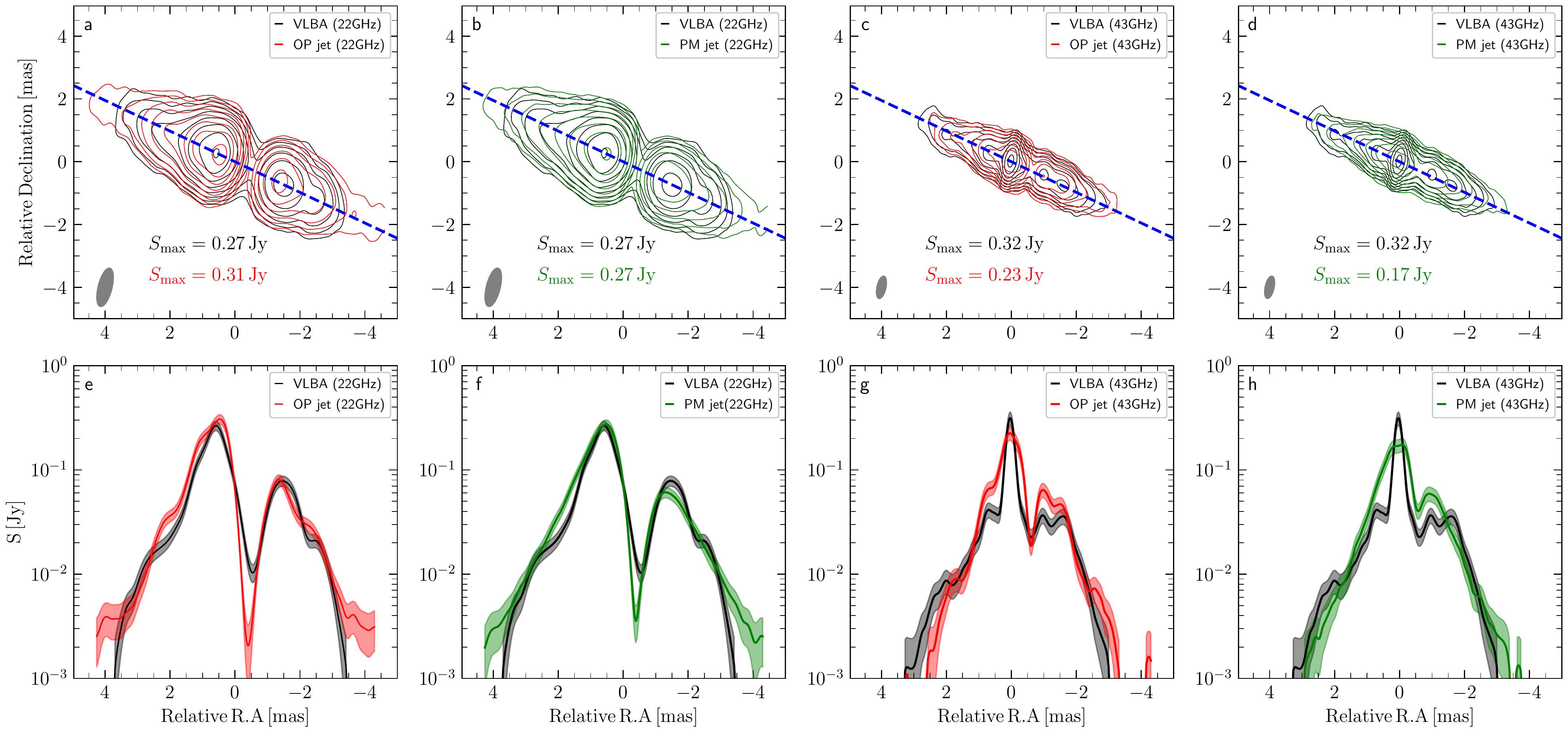}} 
\caption{Results of the non-linear optimisation for NGC\,1052 for 22\,GHz (columns 1 and 2)
and for 43\,GHz (columns 3 and 4): Panels a--d show flux density contours for the VLBA
observation (black) and the jet models (OP jets in red and PM jet green) and panels e--h display
the flux density profiles along the blue dashed line in panels a--d.
The black points correspond to the VLBA observations of NGC\,1052, red to the OP jet model
and green to the PM jet.
In panels a-d, the peak flux density, $S_{\rm max}$, is indicated in the middle and the convolving
beam is plotted in lower left corner of each panel.
The lowest flux density contour is drawn at 1\,mJy and increases by a factor 2.}
\label{bestpos}
\end{figure*}
%
%
In Fig.~\ref{bestpos} we compare the jet structure and the flux density along the jet axes between the
models and observations of NGC\,1052.
Both jet models OP and PM jet are in good agreement with the observed jet structure
(panels a-d in Fig.~\ref{bestpos})
and a more detailed view of the distribution of the flux density along the jet axis is provided in panels e-h.
As mentioned in Sect.~\ref{modelsetup}, OP jets generate local maxima in pressure and density,
i.e., recollimation shocks, while PM jets show a monotonic decrease in pressure and density.
This behaviour is clearly visible in the flux density profiles along the jet axis (panels e-h).
In the 22\,GHz flux density profiles (panels e and g) the OP jet shows a plateau like feature around
$\pm2$ mas, whereas the flux density profile of the PM jet is continuously decreasing.
With increased resolution, \ie higher observing frequency, recollimation shocks closer to the jet nozzle
can be resolved and additional local flux density maxima appear (see panel f at $\pm1$ mas) in contrast
to the PM jet (see panel h)\\

\begin{figure}[h!]
\resizebox{\hsize}{!}{\includegraphics{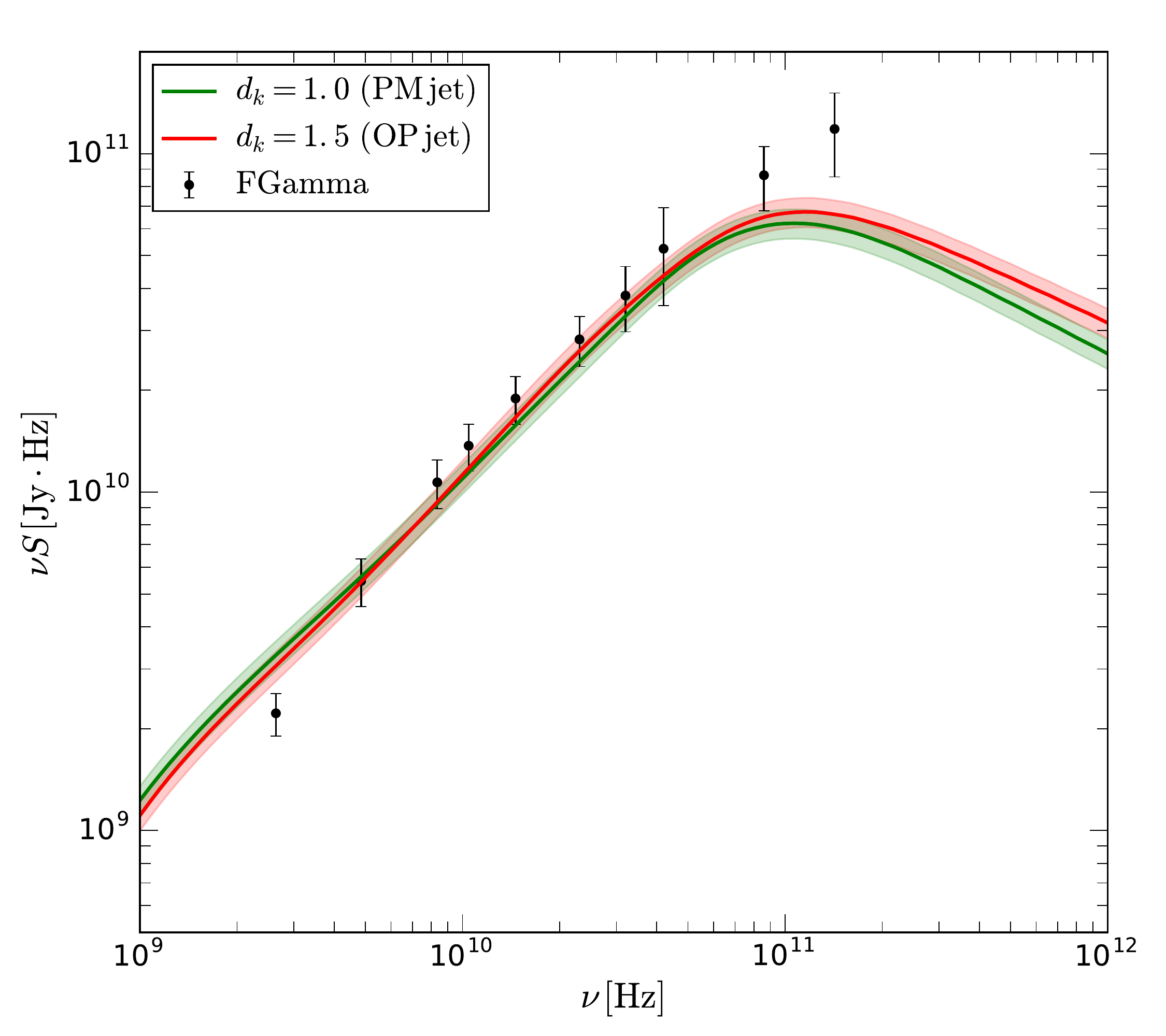}} 
\caption{Broadband radio spectrum of NGC\,1052 including the averaged observations and the 
simulated
OP and PM jet models.
The red and blue shaded regions correspond to $10\%$ flux density variations (see text for details).}
\label{sedfitted} 
\end{figure}

The broadband radio spectrum between $10^9\,\mathrm{Hz}<\nu<10^{12}\,\mathrm{Hz}$
for the different jet models and NGC\,1052 can be seen in Fig.~\ref{sedfitted}.
Typically the radio emission seen by single dish telescopes includes radiation emerging
from large scale structures not observed by VLBI observations.
We take  this observational effect into account by including $10\%$ flux density variations indicated by 
the blue and
green shaded bands around the simulated spectrum.
Both models, OP and PM jets are able to reproduce the observed spectrum,
whereas the OP jet model provides a slightly better fit to the observed spectrum of NGC\,1052
(see $\chi^2$ values in Table \ref{parabest}).
Both models exhibit the trend towards lower high-frequency emission ($\nu>10^{11}\,\mathrm{Hz}$) and
higher low frequency emission ($\nu<5\times10^{9}\,\mathrm{Hz}$).
This behaviour can be attributed to the idealised treatment of the non-thermal particles.
In our simulations we neglect radiative losses and re-acceleration acting on the non-thermal particles.
In Sect.~\ref{discussion} we provide a detailed discussion of the impact of the above mentioned physical
mechanism on the broadband spectrum.\\
%
%
\begin{figure*}[t]
\centering
\resizebox{18cm}{!}{\includegraphics{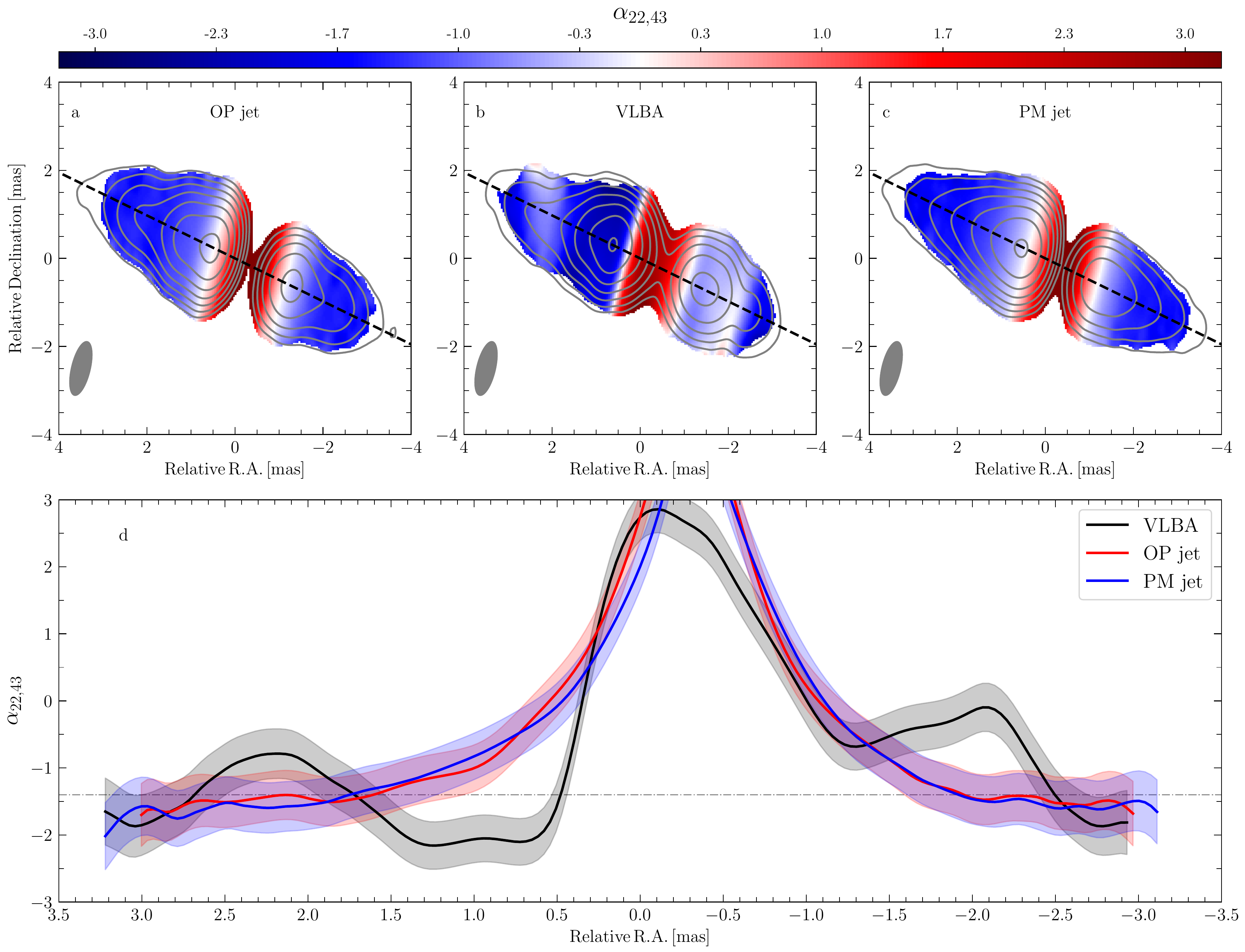}} 
\caption{Spectral index maps for  NGC\,1052.
Panels a - c show: (a) the 2D-distribution of the spectral index for the OP jet model, (b) the VLBA 
observations of NGC\,1052
and (c) the PM jet model.
The variation of the spectral index along the jet axis (black dashed line in panels a-c) can be seen in 
panel d.
In panels a-c, the convolving beam is plotted in the lower left corner of each panel.
The contours correspond to the 22\,GHz flux density distribution and the lowest flux density contour is 
drawn at 1\,mJy and
increases by factors of 2.
The dash-dotted line in panel d corresponds to the optically thin spectral index of the models ($\alpha=1.4$).}
\label{bestposspix}
\end{figure*}
%
%
A glimpse into the acting radiation microphysics in relativistic jets can be obtained via spectral index 
studies
and the computed spectral index between two frequencies $\nu_1$ and $\nu_2$
is related to the energy distribution of the non-thermal particles,
$s$, via $s=-(2\alpha-1)$.
Changes in the spectral index can be attributed to losses including adiabatic (expansion) and radiative
(synchrotron and inverse Compton) \citep[see, \eg][]{Mimica:2009de,2016A&A...588A.101F} and to
re-acceleration of non-thermal particles, \eg internal shocks, shear and magnetic reconnection
\citep[see, \eg][]{2015SSRv..191..519S,2017ApJ...842...39L}.
For the calculation of a spectral index in each pixel of the radio map we convolve both radio images with 
a common beam,
typically the one of the smaller frequency (here 22\,GHz) and align the structure by common
optically thin features in the
jet \citep[see][for details]{2013A&A...557A.105F}.
The computed spectral indices between 22\,GHz and 43\,GHz are presented in Fig.~\ref{bestposspix}.\\
The spectral index computed from the OP and PM model approximates at large distances the theoretical
value of $\alpha=1.4$ (OP jet) and $\alpha=1.45$ (PM jet). The small variation in the spectral index for the models
is an artefact of the image reconstruction algorithm and the convolution with observing beam (see Appendix for details).
The largest difference between our numerical models and the VLBA observations occurs at a distance of $x=\pm2\,\mathrm{mas}$.
These locations coincides in the case of the OP jet with the position of a recollimation shock. As shown in \citet{2016A&A...588A.101F} 
the spectral index at the location of recollimation shocks can be increased or inverted if radiative losses are taken into account. However due to computational limitations we excluded radiative cooling in our emission calculations\footnote{Including radiative
losses similar to \citet{Mimica:2009de,2016A&A...588A.101F} requires the injection and propagation of 
Lagrangian particles which is currently numerically too expensive within our modelling algorithm.}. The central region of NGC\,1052 is dominated by absorption due the obscuring torus, which is reflected by large spectral index values exceeding the typical value for synchrotron self-absorption of $\alpha>2.5$.
With increasing distance from the centre, the spectral index is decreasing (less absorption due the obscuring torus) and approaches the
already mentioned value of $\alpha=1.4$ (OP jet) and $\alpha=1.45$ (PM jet).

\begin{figure}[h!]
\resizebox{\hsize}{!}{\includegraphics{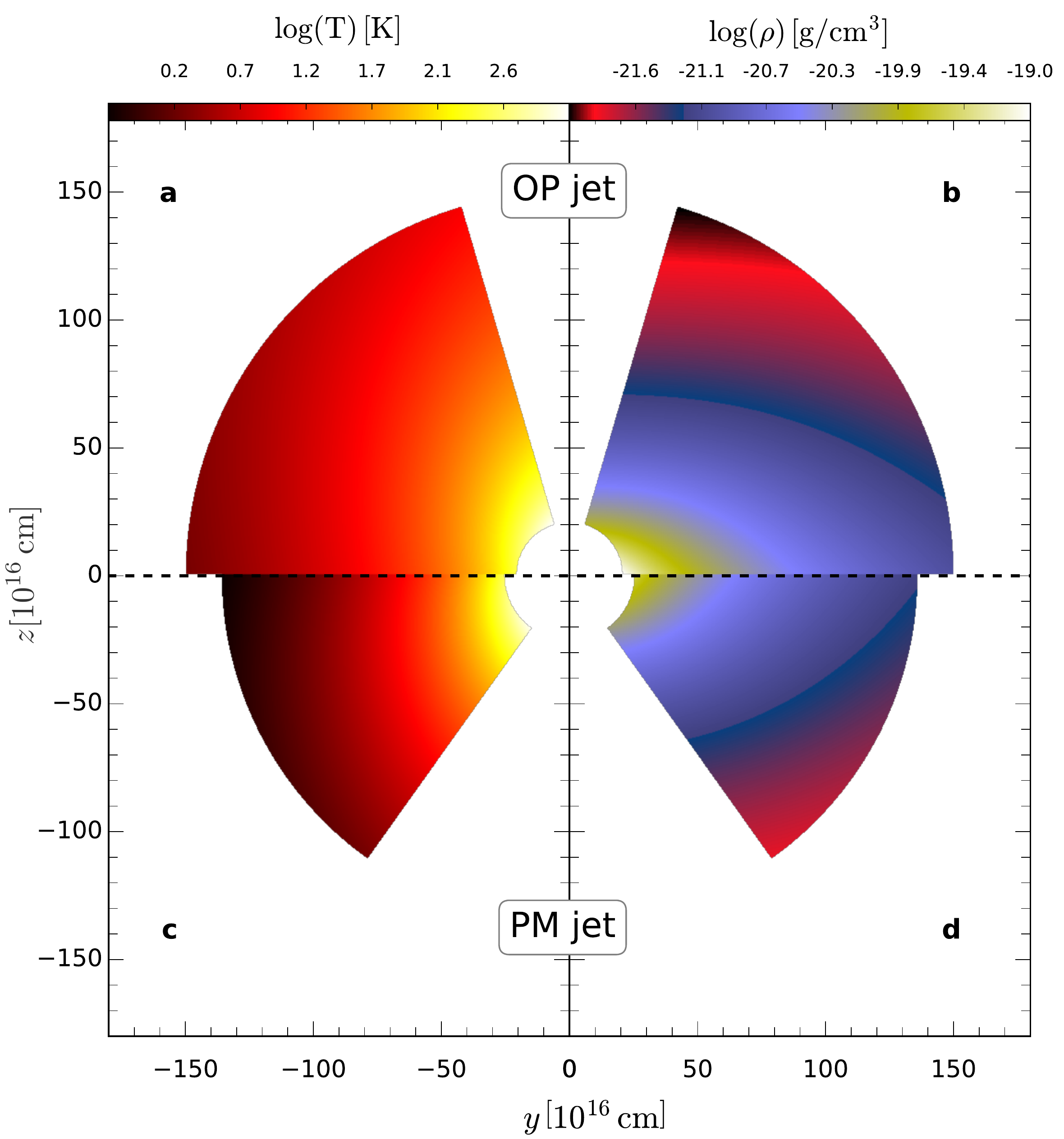}} 
\caption{Distribution of the temperature (panels a and c) and the density (panels b and d) in the torus for 
the
OP jet (top) and the PM jet (bottom).}
\label{torusstruct} 
\end{figure}

In Fig.~\ref{torusstruct} we present the temperature and density distribution within the torus.
The top panels (a and b) show the torus for the OP jet model and the bottom panels (c and d) for the PM 
jet model.
The inner and outer radii for both models are very similar $R_\mathrm{in}=0.2\times10^{18}\,
\mathrm{cm}$ and
$R_\mathrm{out} =1.5\times10^{18}\,\mathrm{cm}$ ($0.5\,\mathrm{pc}$) which is within the estimates
$0.1\,\mathrm{pc}\leqslant R_{\mathrm{torus}}\leqslant 0.7\,\mathrm{pc}$ reported by \citet{2003PASA...20..134K}.
The torus in the OP jet model has a larger opening angle and less steep temperature gradient than the 
torus of the
PM jet (see Table \ref{parabest}). \\
A measurable quantity of the obscuring torus is the number density, which can be computed by 
integrating the density
profile along a line sight.
In our case the density of the torus is modelled via:
\begin{equation}
\rho=\rho \left( R_{\rm in}\right) \left(\frac{r}{R_{\rm in}}
\right)^{-k_\rho}e^{-l_\rho\left| \cos{\Theta}\right|} \,,
\label{rhotorus}
\end{equation}
where the exponents $k_\rho$ and $l_\rho$ model the decrease in density in radial and $\Theta$ 
directions.
Using the values obtained from the non-linear optimisation (see Table \ref{parabest}) we calculate a 
number density of:
\begin{eqnarray}
 N_\mathrm{H}&=&0.7\times10^{22}\,\mathrm{cm^{-2}} \qquad \mathrm{OP\,\, jet} \,, \\
 N_\mathrm{H}&=&1.0\times10^{22}\,\mathrm{cm^{-2}} \qquad \mathrm{PM\,\, jet} \,.
\end{eqnarray}
Both values are in agreement with number density in NGC\,1052 derived from X-ray observations of
$0.6\times10^{22}\,\mathrm{cm^{-2}}\leqslant N_{\rm H,obs}\leqslant 0.8\times10^{22}\,\mathrm{cm^{-2}}
$
\citep{2004A&A...420..467K}.
\section{Discussion}
\label{discussion}
\subsection{Multi-frequency VLBA observations and core-shifts}
\begin{figure*}[t]
\resizebox{\hsize}{!}{\includegraphics{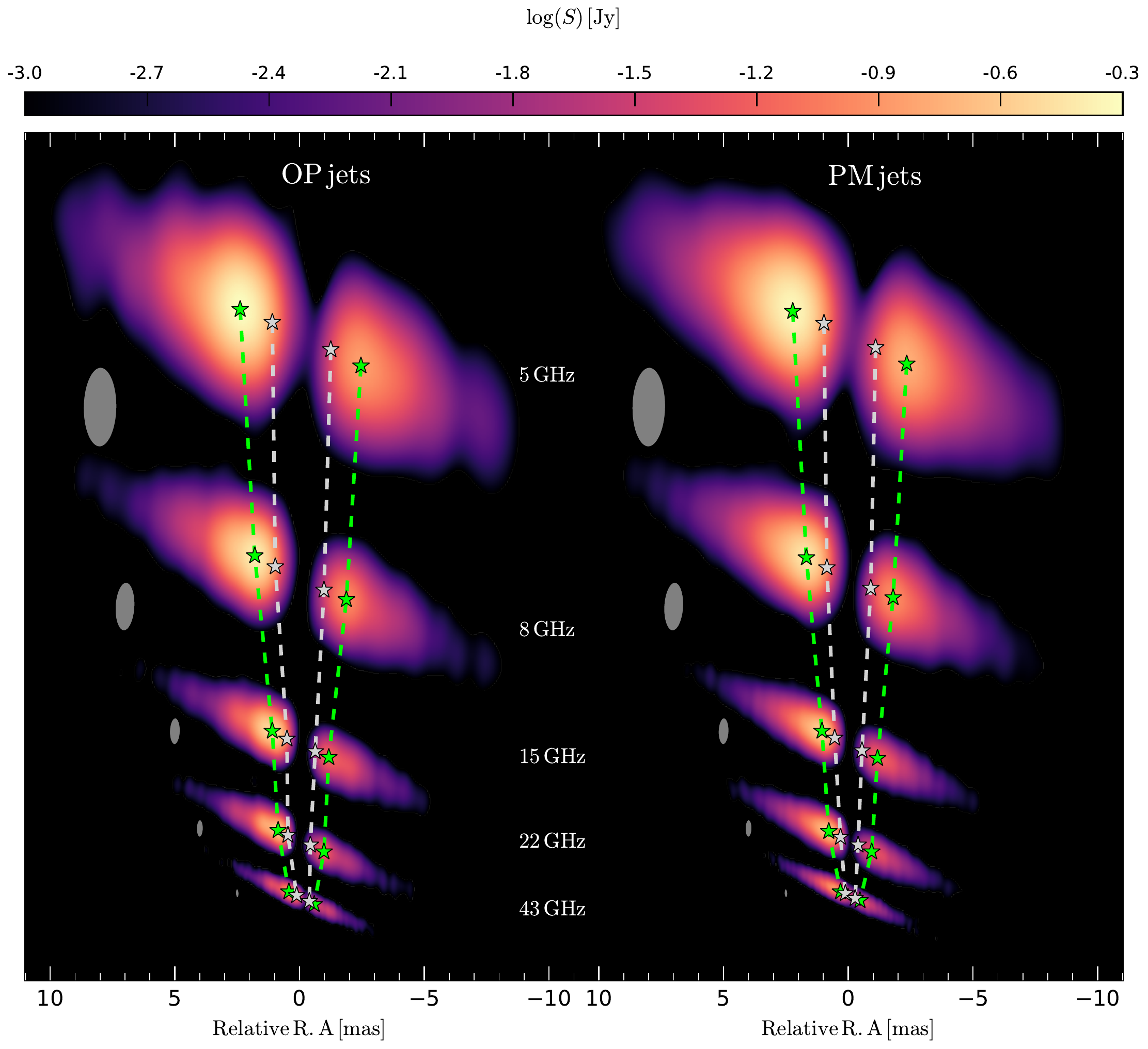}}
\caption{Synthetic multi-frequency VLBA images from 5\,GHz towards 43\,GHz computed for the OP jet
model (left) and the PM jet model (right).
The images are aligned by the centre of the emission gap and in all images the lowest flux density 
plotted is 1\,mJy.
The convolving beam is plotted next to the radio images.
The green stars mark the position of the flux density maximum and the white stars correspond to the 
position where
the flux density reaches 20\% of the peak flux density for the first time.
The dashed lines trace the position of the aforementioned locations through the multi-frequency radio 
images.}
\label{multifreqVLBA}
\end{figure*}

\begin{figure}[h!]
\resizebox{\hsize}{!}{\includegraphics{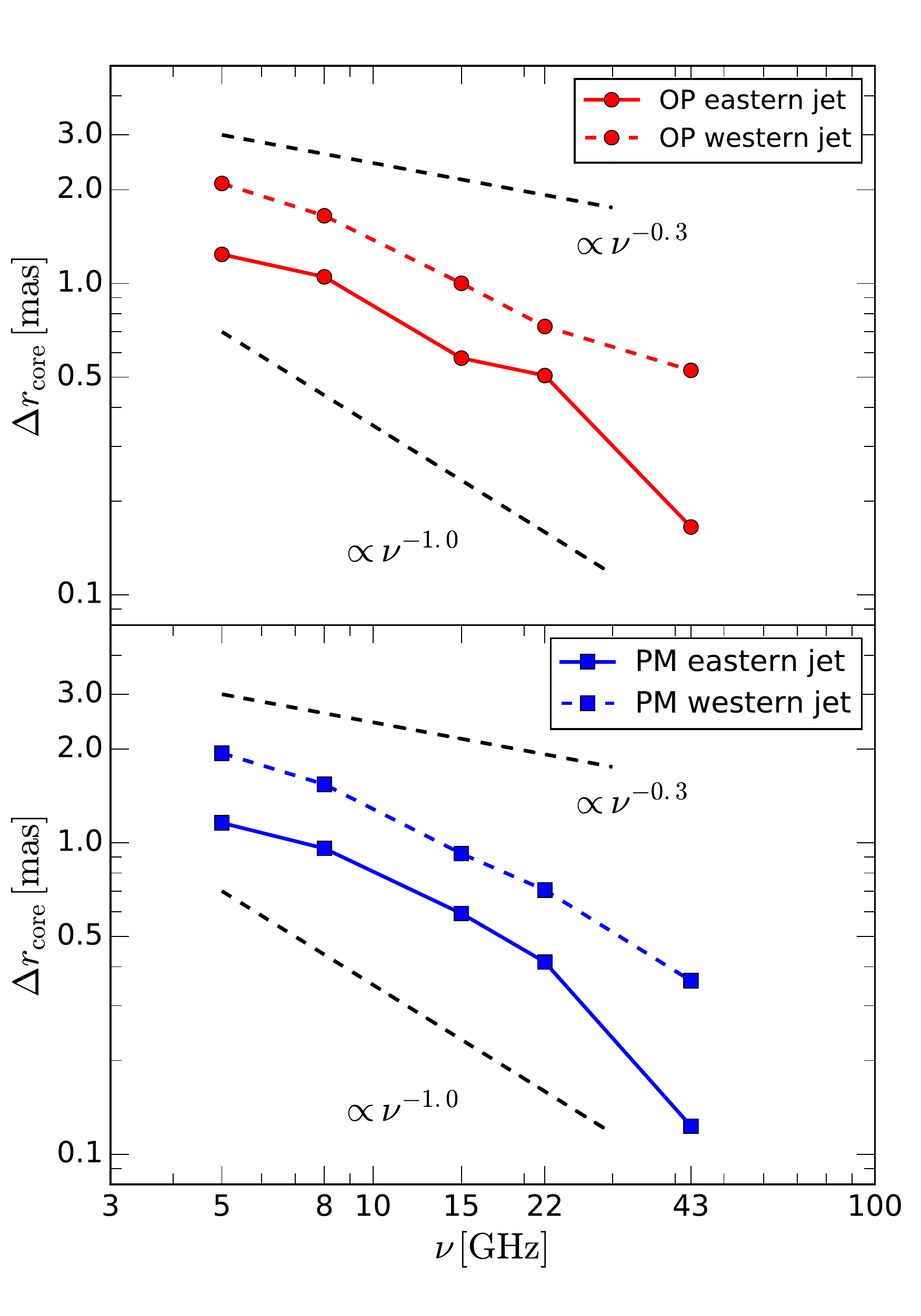}} 
\caption{Results of the core-shift analysis for the OP jet (top panel) and PM jet (bottom panel).
Solid lines correspond to the eastern (left) jet, dashed lines to the western (right) jet.}
\label{coreshift} 
\end{figure}

The inclusion of the broadband radio spectrum (see Fig.~\ref{sedfitted}) in our non-linear optimisation 
process
enables us to compute various synthetic images in addition to 22\,GHz and 43\,GHz.
Therefore, we can model the multi-frequency behaviour of the source and compare the obtained results 
to the
observations of NGC\,1052.
The most remarking feature in the VLBA observations of NGC\,1052 is the emission gap between the 
eastern (left)
and western (right) jet, which is shrinking with increasing frequency 
\citep[see Fig.~1 in][]{2004A&A...426..481K}.
Thus, a valid model of NGC\,1052 must reproduce this behaviour.
In Fig.~\ref{multifreqVLBA} we present our synthetic multi-frequency images for both jet models, OP jet 
and PM jet.
Since the absolute position of the jets is lost during the image reconstruction, we align the jets by the 
centre of the
emission gap.
The emission gap is clearly visible at lower frequencies and the distance between the jets is decreasing 
with
increasing frequency.
As mentioned in Sect.~\ref{results}, the gap between the jets is produced by the combined absorption of 
thermal
particles in the torus and the non-thermal particles in the jet, where the thermal particles provide the 
major
contribution to the appearance of the emission gap at lower frequencies. \\
%
A more qualitative discussion on which particle distribution is dominating the absorption process at 
different frequencies
can be obtained by means of the core-shift.
The radio core of a relativistic jet is usually defined as the $\tau=1$ surface, \ie the onset of the jet.
The optical depth $\tau$ is computed from the sum of absorption coefficients for the thermal distribution, $\alpha_{th}$, and non-thermal particle distribution, $\alpha_{nt}$,  along the line of 
sight:
\begin{equation}
\tau=\int \left(\alpha_{th}+\alpha_{nt}\right) \, ds \,.
\end{equation}
Since the absorption coefficient depends on the physical conditions in the jet, (\eg density, temperature 
and magnetic field)
and on frequency, the shift in the core position can be used to probe the radiation micro-physics 
\citep{1998A&A...330...79L}.\\ 
In the analysis of VLBI data, the jet is typically modelled via several gaussian components representing 
the observed brightness
distribution, and the innermost component is selected as the core.
However, such a detailed modelling of our synthetic radio images is beyond the scope of this work.
Based on detailed modelling of the NGC\,1052 observations by \citet{2004A&A...426..481K} we define 
the observed onset of the jet
as the innermost location where the flux density reaches $\sim20\%$ of the peak flux density in each 
jet\footnote{This value is
derived from the observed peak flux density and flux density of the innermost gaussian component for 
different frequencies using
values given in Table 1 of \citet{2004A&A...426..481K}.}
In Fig.~\ref{results} the green stars correspond to the flux density maximum in the eastern and western 
jet, and the white stars mark
the onset of the jets using the method described above. 
The frequency-dependent variation of the core-shift with respect to the centre of the torus is presented in 
Fig.~\ref{coreshift}.
At lower frequencies the distance between the jets decreases as $\Delta r\propto \nu^{-0.3}$ and 
continuously steepens
towards $\Delta r\propto \nu^{-2}$.\\
%
This change in the slope is an indication of a change in the radiation process responsible for the 
absorption.
On larger scales, which are probed by lower frequencies, the opacity is dominated by the thermal 
particle distribution in the torus.
Therefore, the absorption coefficient (in the Rayleigh-Jeans regime $h\nu \ll kT$)  can be written as:
\begin{equation}
\alpha_{\rm th,RT}\propto T^{-3/2}\rho^2\nu^{-2} \,.
\end{equation}
Inserting the temperature and density profiles used for the obscuring torus
(and omitting the angular dependence for simplicity):
\begin{eqnarray}
T\propto r^{-l} \,,\\
\rho\propto r^{-k} \,,
\end{eqnarray}
the variation of the core position with frequency can now be written as:
\begin{equation}
r_{\rm core,th}\propto \nu^{-1/(0.75 l + k-0.5)} \,.
\label{corethermal}
\end{equation}
Inserting the values for $k$ and $l$ obtained from the non-linear optimisation in Eq.~\ref{corethermal} 
leads to:
\begin{eqnarray}
 r_{\rm core,th}\propto \nu^{-0.27} \qquad \mathrm{OP\,jet} \label{rcoreOPth} \,,\\
 r_{\rm core,th}\propto \nu^{-0.22} \qquad \mathrm{PM\,jet} \,.
 \label{rcorePMth}
\end{eqnarray}
With increasing frequency, the absorption due to the thermal particle distribution is decreasing and the 
non-thermal
particles start to dominate the opacity.
Following \citet{1998A&A...330...79L} the core position can be written as:
\begin{equation}
r_{\rm core,nt} \propto \nu^{-(5-2\alpha)/\left[\left(2\alpha -3\right)b-2n-2\right]} \,,
\label{corenonthermal}
\end{equation}
where $\alpha$ is the spectral index and $b$ and $n$ are the exponents of the evolution of the magnetic 
field,
$B\propto r^{-b}$, and the density of the non-thermal particles, $n_0\propto r^{-n}$, with distance.
In the case of equipartition (kinetic energy density equals magnetic energy density), 
Eq.~\ref{corenonthermal}
leads to: $r_{\rm core,nt}\propto \nu^{-1}$ using $m=1$ and $n=2$.
\begin{figure}[h!]
\resizebox{\hsize}{!}{\includegraphics{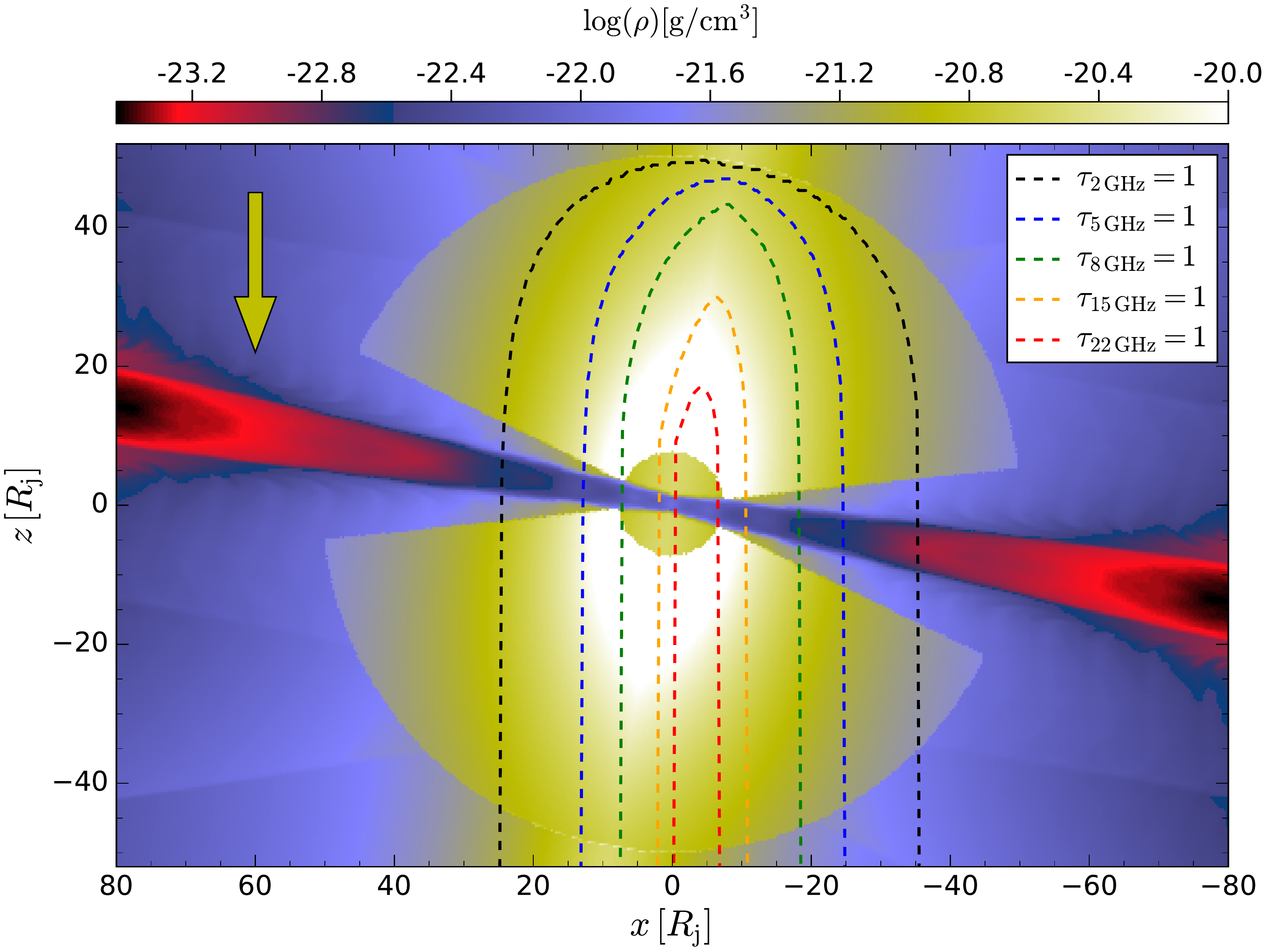}} 
\caption{Distribution of the rest mass density for the OP jet in the z-x plane at y=0.
The dashed lines trace the $\tau=1$ surfaces for 2\,GHz, 5\,GHz, 8\,GHz, 15\,GHz and 22\,GHz.
The yellow arrow indicates the direction of the ray-tracing and regions of the jet within the dashed area 
are obscured
by the torus and not visible in the corresponding radio images.}
\label{tausurface} 
\end{figure}
If the non-thermal particle distribution dominates the absorption along a line of sight, we can derive 
estimates for
the evolution of the magnetic field and the non-thermal particles using Eqns.~\ref{Bcal} and \ref{nnorm}.
Both equations depend on the pressure: $B\propto p^{1/2}$ and $n_0\propto p$.
The evolution of the pressure is given by eq.~\ref{pamb}, which is correct for PM jets and differs only in 
the position
of the recollimation shocks in the case of OP jets.
The evolution of the pressure can be divided into two different regimes: $p\propto r^{-2}$ for $r >r_c$ 
and
$p \propto r^{-4/3}$ for $r\leqslant r_c$.
Inserting the values reported in Table \ref{parabest} into Eq.~\ref{corenonthermal} leads to the following
core-shift behaviour in the case of non-thermal particles dominating the absorption:
\begin{eqnarray}
r_{\rm core,nt} &\propto& \nu^{-1} \qquad \,\,\, r >r_c \label{coreshiftrc} \,, \\
r_{\rm core,nt} &\propto& \nu^{-2.7} \qquad r\leqslant r_{\rm c} \,.
\label{coreshiftr}
\end{eqnarray}
Using the developed estimates for the variation of the core position with frequency,
Eqns.~\ref{rcoreOPth}--\ref{coreshiftr}, we can explain the core-shift behaviour
presented in Fig.~\ref{coreshift}.
At lower frequencies, here between 5\,GHz and 8\,GHz, the thermal particle distribution
in the torus provides the major contribution to the absorption along the line of sight
which leads to a slope of $-0.3$.
This value is in agreement with the derived values of $-0.27$ and $-0.22$, respectively.
As the frequency increases, the torus becomes more transparent and the absorption due
to the non-thermal particles in the jet starts contributing to the opacity.
Therefore, we obtain for $8\,\mathrm{GHz}<\nu<22\,\mathrm{GHz}$ a steepening of the
core-shift from $\propto \nu^{-0.3}$ to $\propto\nu^{-1}$.
At frequencies $\nu>22$\,GHz we probe the regions close to the jet nozzle and
as mentioned above, there is a change in the pressure gradient. Therefore we expect
a strong steepening of the core-shift in this region.
This behaviour is clearly visible in Fig.~\ref{coreshift} for $\nu>22\,$GHz for the eastern jets.\\
Due to the geometry of the jet-torus system relative to the observer, the path of a light ray
through the obscuring torus is longer for a ray emerging from the western jet than for a ray
leaving the eastern jet.
Thus, the thermal absorption for a light ray from the western jet is larger than for its eastern
counterpart.
Therefore, we expect the steeping of the core-shift to be shifted to higher frequencies in
the case of the western jet.
This effect can be seen in the western jets for both jet models (see Fig.~\ref{coreshift}).
To illustrate the frequency-dependent position of the $\tau=1$-surface and its impact on the
visible regions in the radio images we plot the opacity for four different frequencies on top of
the z-x slice of the rest-mass density for the OP jet model (see Fig.~\ref{tausurface}).

\subsection{Over-pressured vs. pressure-matched jets}
So far both models, OP jet and PM jet, provide very similar results and can successfully reproduce
the observations of NGC\,1052.
It is therefore difficult to distinguish both models based on current observations.
A major difference between OP jets and PM jets is the generation of recollimation shocks.
However, at low frequencies the torus can mimic a recollimation shock, \ie a local flux density maxima,
by a drop in opacity (see for example the flux density peak -1\,mas in panel h of Fig.~\ref{bestpos}).
A way out of this dilemma can be provided by $\mu$as resolution VLBI observations.
This high resolution can be achieved in two ways: increasing the observing frequency and/or increasing
the baselines.
The space-based radio antenna of the RadioAstron satellite operates at 1.6\,GHz, 5\,GHz and 22\,GHz,
and extends the projected baseline up to 10 Earth radii \citep{2013ARep...57..153K}.
In addition to the space-based antenna, there are two more VLBI experiments providing $\mu$as -
resolution:
The Global Millimetre VLBI Array (GMVA) at 86\,GHz \citep{2008AJ....136..159L} and the
Event Horizon Telescope (EHT) at 230\,GHz \citep{2017NatAs...1..646D}.
At 230\,GHz we will observe regions close to the central engine, requiring a general-relativistic treatment
of the MHD and the radiative transport, which will be addressed in a future work.
Therefore, we focus in this paper on synthetic radio images as observed by RadioAstron at 22\,GHz and 
the
GMVA at 86\,GHz.
In the Appendix we provide details on the observing schedule, the array configuration and the SEFDs for 
the two arrays.
For the reconstruction of the synthetic radio images we apply a maximum entropy method \texttt{MEM}
provided by the \texttt{EHTim} package, since the \texttt{CLEAN} algorithm tends to produce patchy
structure for smooth flux density distributions.\\

\begin{figure}[h!]
\resizebox{\hsize}{!}{\includegraphics{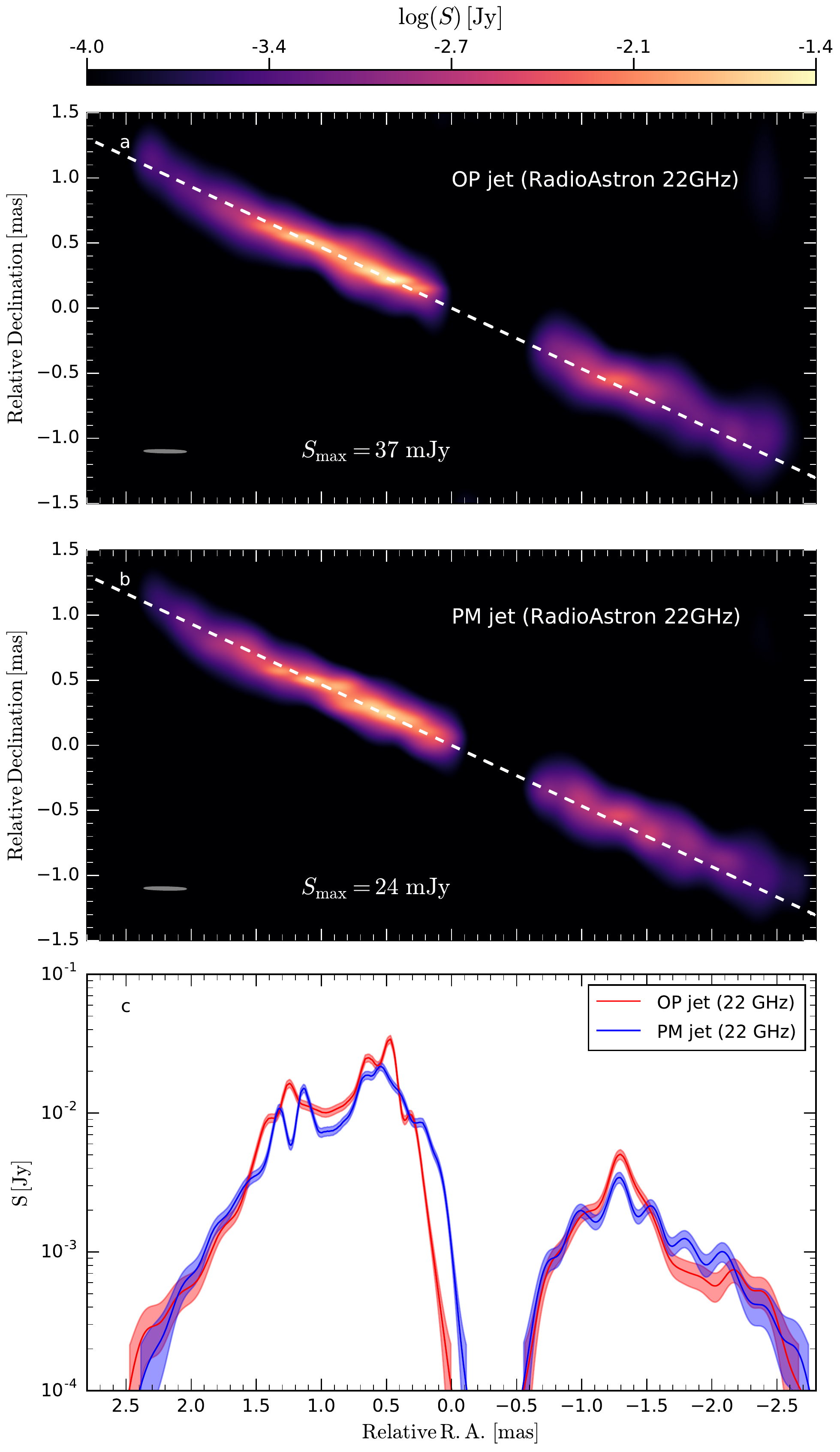}} 
\caption{Synthetic radio images for NGC\,1052 as seen by the RadioAstron satellite (panels a and b) at 
22\,GHz.
The OP jet is presented in panel a and PM jet in panel b.
The convolving beam, $160\,{\rm \mu as} \times 20\,{\rm \mu as}$, is plotted in the lower right corner.
The flux density along the jet axis, white dashed line in panels a and b, is shown in panel c.
See text and Appendix for details on the observation schedule and the array settings.}
\label{RadioAimage} 
\end{figure}

The results of the $\mu$as resolution imaging for our jet models are presented in 
Fig.~\ref{RadioAimage}.
The 22\,GHz RadioAstron images show a clear emission gap between the eastern and western jet,
similar to the 22\,GHz VLBA images (see Fig.~\ref{multifreqVLBA}).
The synthetic radio images from both models show a similar flux density distribution and more details 
can
be seen in the flux density along the jet axis (panel c in Fig.~\ref{RadioAimage}).
The variation of the flux density along the jet axis is comparable in both models and therefore it is not 
possible
to distinguish both models based on RadioAstron images. 

%

At 86\,GHz the torus is optically thin and a clear view to the central region is obtained, \ie no absorption 
due
the obscuring torus is seen.
The GMVA observations provide, for the first time, a clear detectable difference between the OP jet and 
PM jet:
there is a local flux density maximum at $\pm0.5$mas which is not seen in the PM jet.
Given the most recent GMVA observations of NGC\,1052 there are two bright features next to core at 
roughly
$\pm0.4$mas \citep{2016A&A...593A..47B} which could be interpreted as the recollimation shocks.\\
Based on this result, together with the variation in the spectral index (see Fig.~\ref{bestposspix}), we 
favour
the over-pressured jet scenario as the most likely configuration of NGC\,1052.

\begin{figure}[h!]
\resizebox{\hsize}{!}{\includegraphics{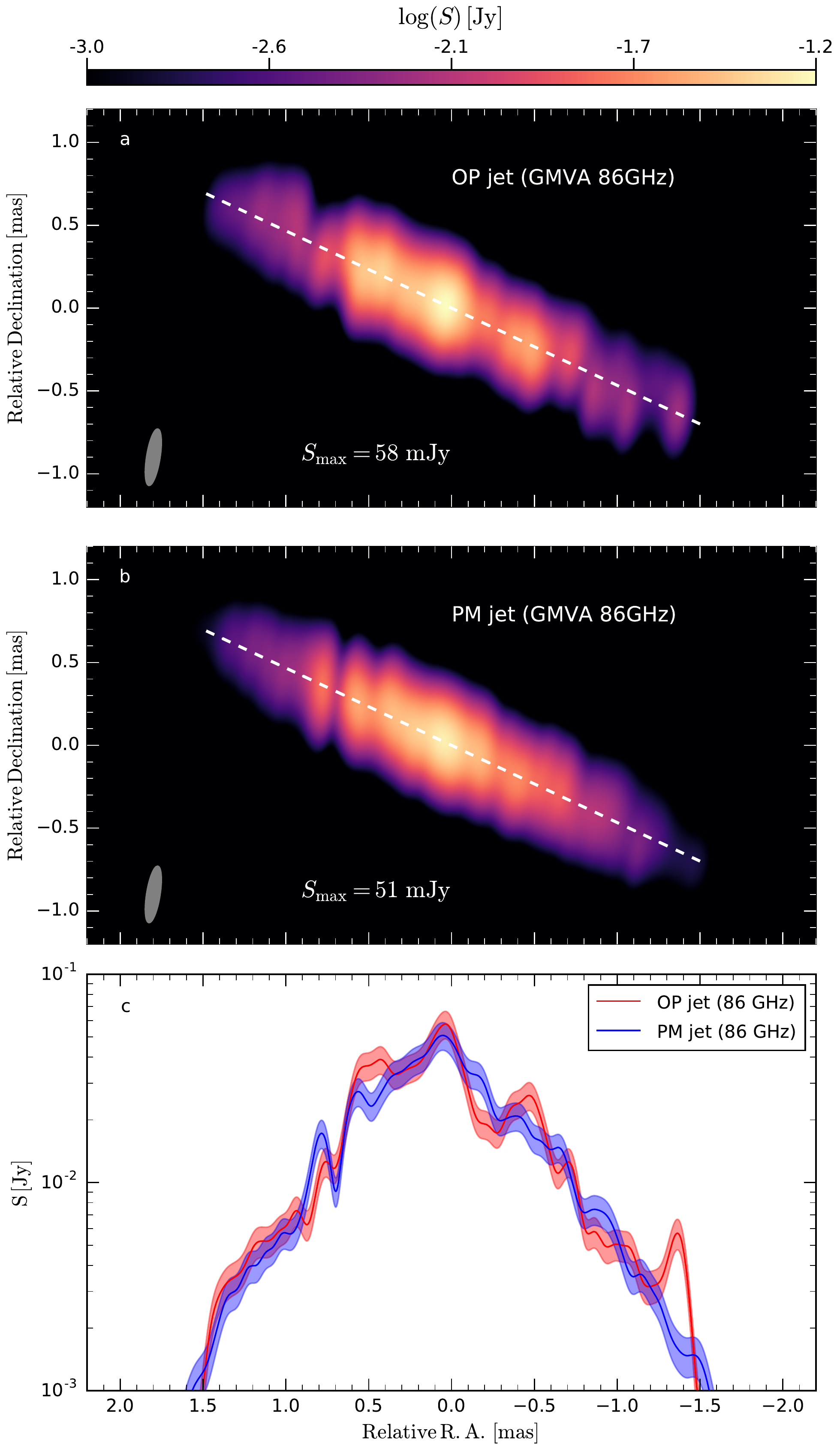}} 
\caption{Synthetic radio images for NGC\,1052 as seen by the GMVA (panels a and b) at 86\,GHz.
The OP jet is presented in panel a and the PM jet in panel b.
The convolving beam, $345\,{\rm \mu as} \times 82\,{\rm \mu as}$, is plotted in the lower right corner.
The flux density along the jet axis, white dashed line in panels a and b is shown in panel c.
See text and Appendix for details of the observation schedule and the array settings.}
\label{GMVAimage} 
\end{figure}
%
%
%
\subsection{Refining the SRHD model}
Based on the discussion in the previous section we conclude that NGC\,1052 is best modelled by an 
over-pressured jet.
We can further improve our model for NGC\,1052 if we modify the underlying SRHD models with respect 
to the
pressure-mismatch, $d_k$, and the gradients in the ambient medium.
Therefore, we produce several SRHD simulations changing $d_k$ in 0.1 steps and $z_c$ in steps of 2.
We inject these models in our pipeline and use the values reported in Table \ref{parabest} as initial 
positions for the
parameter search.
To avoid biasing effects during the optimisation we add some random scatter on the initial position.
This approach has the advantage that computational efforts for the refined study are lower than for an 
optimisation
starting from random positions in the parameter space (as performed in Sect.~\ref{results})

\begin{figure*}[t!]
\centering
\resizebox{18cm}{!}{\includegraphics{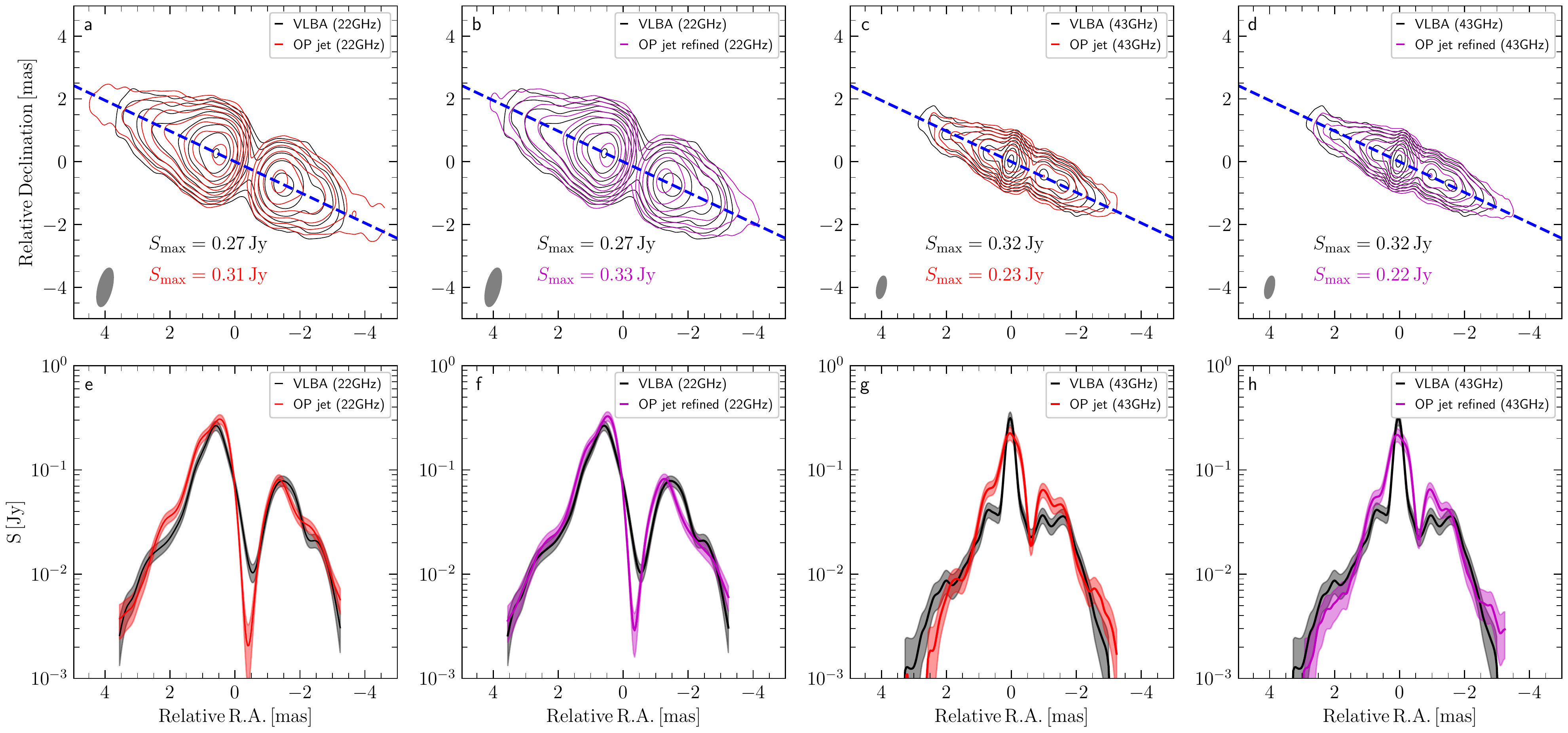}} 
\caption{Same as Fig.~\ref{bestpos} including the refined OP jet model.
See text for details.}
\label{bestposref}
\end{figure*}

The result of the refinement of the underlying SRHD simulation is presented in Fig.~\ref{bestposref} and 
the obtained
values are presented in Table \ref{refinedparabest}.
The main difference between the initial model and the refined one is the core-radius of the ambient 
medium,
which reduced from $z_c=10\,R_{\rm j}$ to $z_c=8\,R_{\rm j}$, while $d_k=1.5$ is not changed.
Reducing the core-radius, $z_{\rm c}$, induces an earlier transition from $p_a(z)\propto z^{-1.3}$ to
$p_a(z)\propto z^{-2}$ (see Eq.~\ref{pamb}).
Due to the steeper decrease in the ambient pressure, the position where a transversal equilibrium 
between the pressure
in jet and the ambient medium is established is shifted downstream.
Therefore, the recollimation shocks will be formed at a larger distance from the jet nozzle as compared 
to the jet
embedded in an ambient medium with a larger core size.
This effect can be seen in flux density cuts along the jet axis in panels e--h in Fig.~\ref{bestposref}.
The local flux density maximum at 2\,mas is better approximated by the refined model than by the initial 
one.
In addition, the innermost flux density peaks at $\sim1$\,mas are also better fitted by the refined model.
%

%
\begin{table}[t!]
\setlength{\tabcolsep}{4pt}
\caption{Best fit parameters obtained from OP jet and OP jet refined model}
\label{refinedparabest}
\centering
\begin{tabular}{@{}l l l@{}}
\hline\hline
Symbol  & OP jets &  OP jets (refined)\\
\hline
\multicolumn{3}{c}{\textit{\underline{scaling parameter}}}\\
$d_{\rm k}$		&\multicolumn{2}{c}{1.5}\\
$R_{\rm j}$  		&\multicolumn{2}{c}{$3\times10^{16}$ cm} \\
$z_c$			&$10\,R_{\rm j}$ & $8\,R_{\rm j}$  \\
$n$				&\multicolumn{2}{c}{1.5}\\
$m$				&\multicolumn{2}{c}{2}\\
$z$				&\multicolumn{2}{c}{0.005}\\
$v_{\rm j}$		&\multicolumn{2}{c}{$0.5\,\mathrm{c}$}\\
$\hat{\gamma}$	&\multicolumn{2}{c}{13/9}\\ 
$\rho_{\rm a} $		&$3.0\times10^{-21}\,\mathrm{g \, cm^{-3}}$	    &$5.6\times10^{-21}\,
\mathrm{g \, cm^{-3}}$ \\
\hline \hline
\multicolumn{3}{c}{\textit{\underline{emission parameter}}}\\
$\epsilon_{\rm B}$			 &0.39 	&0.10 \\
$\epsilon_{\rm e}$ 		        &0.27		&0.32 \\
$\zeta_{\rm e}$ 		       &0.35        &0.43 \\
$\epsilon_\gamma$		 &1000		& 1000 \\
$s$						 &3.8		& 3.5\\
$\vartheta$				 &80$^\circ$ 	& 80$^\circ$ \\
\hline \hline
\multicolumn{3}{c}{\textit{\underline{torus parameter}}}\\
$L_{\rm AGN}$		&\multicolumn{2}{c}{$1\times10^{43}\,\mathrm{erg \,s^{-1}}$}\\
$R_{\rm out}$		&$1.4\times10^{18}\,\mathrm{cm}$					& 
$1.5\times10^{18}\,\mathrm{cm}$\\
$\theta$					&73$^\circ$ 	& 54$^\circ$  \\
$\rho\left(R_{\rm in}\right)$	 &$1.1\times10^{-19}\,\mathrm{g \, cm^{-3}}$		& 
$1.0\times10^{-19}\,\mathrm{g \, cm^{-3}}$\\
$T_{\rm sub}$			 &1250 K									& 1170 K\\
$k$, $l$				 &2.4 (2.3), 2.5 (1.5)					 			         & 1.8 (2.8), 
2.9 (1.8) \\
$N_H$		& $0.7\times10^{22}\,\mathrm{cm^{-2}}$ & $1.1\times10^{22}\,\mathrm{cm^{-2}}$ \\
\hline\hline
\\[-1em]
\multicolumn{3}{c}{\textit{\underline{$\chi^2$ results}}}\\
\\[-1em]
$\chi^2_{\rm 22\,GHz}$ & 4.72 & 4.29 \\ 
\\[-1em]
$\chi^2_{\rm 43\,GHz}$ & 7.30 & 6.90 \\
\\[-1em]
$\chi^2_{\rm SED}$     & 1.28 & 1.48 \\
\\[-1em]
\hline \hline
\\[-1em]
\multicolumn{3}{c}{\textit{\underline{image metrics}}}\\
\\[-1em]
$\mathrm{cc}_{\rm 22\,GHz}$ & 0.98 & 0.98 \\ 
\\[-1em]
$\mathrm{cc}_{\rm 43\,GHz}$ & 0.91 & 0.91 \\
\\[-1em]
$\mathrm{DSSIM}_{\rm 22\,GHz}$     & 0.23 & 0.23 \\
\\[-1em]
$\mathrm{DSSIM}_{\rm 43\,GHz}$  &0.10   & 0.10 \\
\\[-1em]
\hline \hline
\end{tabular}
\end{table}
\subsection{Limitations of the model}
Our current model is able to successfully reproduce several features of the VLBA observations,
including the extent of the torus, the number density within the torus, the frequency-dependent
emission gap between eastern and western jets, and the distribution of the flux density and spectral 
index.
However, some details of the NGC\,1052 observations cannot be reproduced with high accuracy.
The observed flux density evolution along the jet axis in the western jet is decreasing faster than
in the OP and PM models, while flux density evolution in the eastern jet is very well-reproduced.
Since our underlying SRHD jet models are symmetric, this could be an indication of:
(i) asymmetries in the ambient medium and/or (ii) asymmetries in the jet launching. 
These limitations could be addressed by future 3D simulations embedded in a slightly asymmetric
ambient medium.\\

Since we use SRHD simulations to model the jets in NGC\,1052, the magnetic field is not 
evolved as an independent parameter and we compute the magnetic field from the 
equipartition pressure (see Eq.~\ref{Bcal}).
Therefore, we restrict ourself to $\epsilon_{\rm b}<0.5$ during the modelling and optimisation process.
Given a viewing angle $\vartheta>80^\circ$, no asymmetries are expected even if the
dominating component of the field is toroidal. From a dynamical point of view, a
strong toroidal magnetic field could reduce the distance between the
recollimation shocks for a given jet overpressure \citep[see, \eg][]{2015ApJ...809...38M,2016ApJ...831..163M}, thus allowing 
for a larger overpressure factor $d_k$ between the jet and the ambient medium in the jets of NGC~1052.

In our current model we ignore the impact of radiative cooling on the relativistic particles.
Depending on the strength of the magnetic field, radiative losses can lead to a steeping of spectral 
indices and
a shortening of the jets at high frequencies \citep{Mimica:2009de,2016A&A...588A.101F}.
In our case, with $\epsilon_{\rm b}<0.5$ we expect only a small impact of the radiative losses on the 
large scale
structure of the jets.
In addition, the magnetic field is decreasing with distance from the jet nozzle which will further reduce its
influence on the jet structure.
However, at the jet nozzle, at high frequencies \citep[$\nu>86\,\mathrm{GHz}$ in the case of NGC\,1052]
[]{2016A&A...593A..47B}
the magnetic field will become important for both the dynamics and the radiative properties of the jet.
Our current model requires a large value for the spectral slope $s\approx 4$ to model the structure of 
NGC\,1052
(especially the distribution of the spectral index).
Such a large spectral slope can be obtained from a non-thermal particle distribution with $s\sim2.2$,
if radiative losses are taken into account.
Especially between the jet nozzle and the first recollimation shock, relativistic particles with large $
\gamma_e$
will suffer radiative losses which will steepen the particle distribution and lead to large spectral slopes.
However, such a self-similar treatment of the non-thermal particles during the optimisation process is
currently computationally too demanding.

\section{Conclusions and outlook}
In this work we present an end-to-end pipeline for the modelling of relativistic jets using 
state-of-the art SRHD and emission simulations coupled via evolutionary algorithms to 
high resolution radio images.
We use this newly-developed pipeline to model stacked radio images and the broadband
radio spectra of NGC\,1052.
The obtained results, \ie synthetic radio images and broadband spectrum, mimic very well
their observed counterparts and the recovered parameters for the obscuring torus are in
agreement with derived estimates from radio and X-ray observations.
The detailed comparison with available data leads to the conclusion that NGC\,1052 is
best described by an over-pressured jet ($d_k=1.5$) in a decreasing pressure ambient
medium ($p\propto r^{-1.3}$).

In a follow up work we will explore the time evolution of the jet in NGC\,1052 and the 
impact of time-delays (slow-light radiative transfer) on the radio structure using the 
obtained values for the jet and torus as initial parameters.
The obtained jet configuration 
can also be used as a framework for further studies including the improvement of our
radiation model including radiative loss and re-acceleration mechanisms of the non-thermal
particles along the flow.\\

To overcome recent limitations with respect to the magnetisation of the jets we will couple in 
a follow-up work our G(S)RMHD and polarised radiative transfer codes to the presented 
pipeline.
This improvement will allow us to drop the limitations on $\epsilon_b$ and 
furthermore enables us to include polarised observations \ie fraction of polarisation and
rotation measures, in the modelling.
The inclusion of polarisation will provide an additional independent constraint on the magnetic
field strength and its geometry, which will allow us to investigate different magnetic field
configurations.

\begin{acknowledgements}
Support comes from the ERC Synergy Grant ``BlackHoleCam - Imaging the
Event Horizon of Black Holes'' (Grant 610058). ZY acknowledges support from a
Leverhulme Trust Early Career Fellowship. 
M.P. acknowledges partial support from the Spanish MICINN grant AYA-2013-48226-C03-02-P 
and the Generalitat Valenciana grant PROMETEOII/2014/069. 
CMF wants to thank Walter Alef and
Helge Rottmann for useful comments and fruitful discussions on synthetic imaging and 
image reconstruction. This research has made use of data obtained with the 
Very Long Baseline Array (VLBA). The VLBA is an instrument of the National Radio 
Astronomy Observatory, a facility of the National Science Foundation operated under 
cooperative agreement by Associated Universities, Inc. This work has made use of NASA's Astrophysics Data System (ADS)
\end{acknowledgements}

\bibliographystyle{aa}

\newpage

\section*{Appendix}
Here we provide details on the convergence studies performed for both the broadband radio 
spectrum and synthetic images, and on the parameter recovery test.
For the convergence study we use the parameters presented in Table \ref{paraemsyn} and 
increased the numerical resolution in $100^3$ steps.

\subsection*{Image convergence }
In order to quantify the convergence of our computed radio image we follow the approach of 
\citet{2016ApJ...817..173L,2018NatAs...2..585M} and use the structured dissimilarity (DSSIM) index 
\citep[for details see][]{2004ITIP...13..600W}. 

\begin{figure}[h!]
\resizebox{\hsize}{!}{\includegraphics{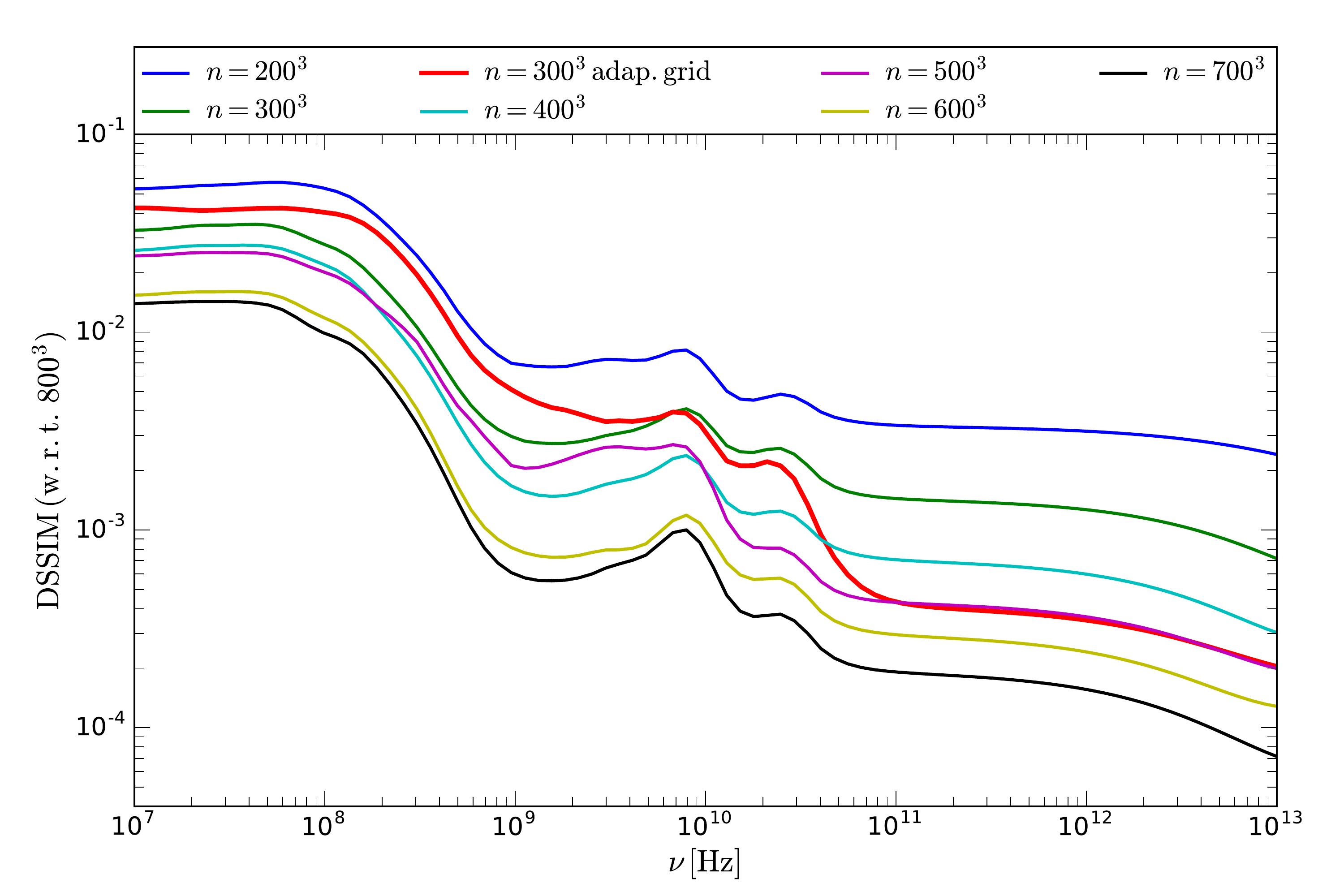}} 
\caption{Convergence test for the synthetic images. The synthetic images produced with 
$800^3$ resolution are used as reference images.}
\label{convimag} 
\end{figure}

Using the definition of the $\mathrm{DSSIM}$, two identical images would have $\mathrm{SSIM}=1$ and $
\mathrm{DSSIM}=0$. For this study we select as reference the images obtained with the 
highest numerical resolution (here $800^3$ cells). The DSSIM is computed between the 
reference image and an image with lower numerical resolution \eg $n=700^3$ cells. For 
each frequency and numerical resolution we obtain a DSSIM value. The result of this study 
can be seen in Fig. \ref{convimag}. The calculated DSSIM varies between 0.05 and 0.0001 
which indicates an overall good agreement between the images.  The DSSIM decreases 
with increased numerical resolution as expected. The local maximum between 
$1\times10^{9}$\,Hz and $5\times10^{10}$\,Hz is due to the torus which leads to higher 
absorption. The red curve indicates the results for the adaptive linear-logarithmic grid. The 
obtained DSSIM for this grid is in the low frequency regime similar to the uniform cartesian 
grid with the same number of cells. However, the strength of this grid becomes obvious at 
higher frequencies where the DSSIM drops below its uniform cartesian counterpart and 
approaches the DSSIM of the $500^3$ grid.  
\subsection*{Spectral convergence}
\begin{figure}[h!]
\resizebox{\hsize}{!}{\includegraphics{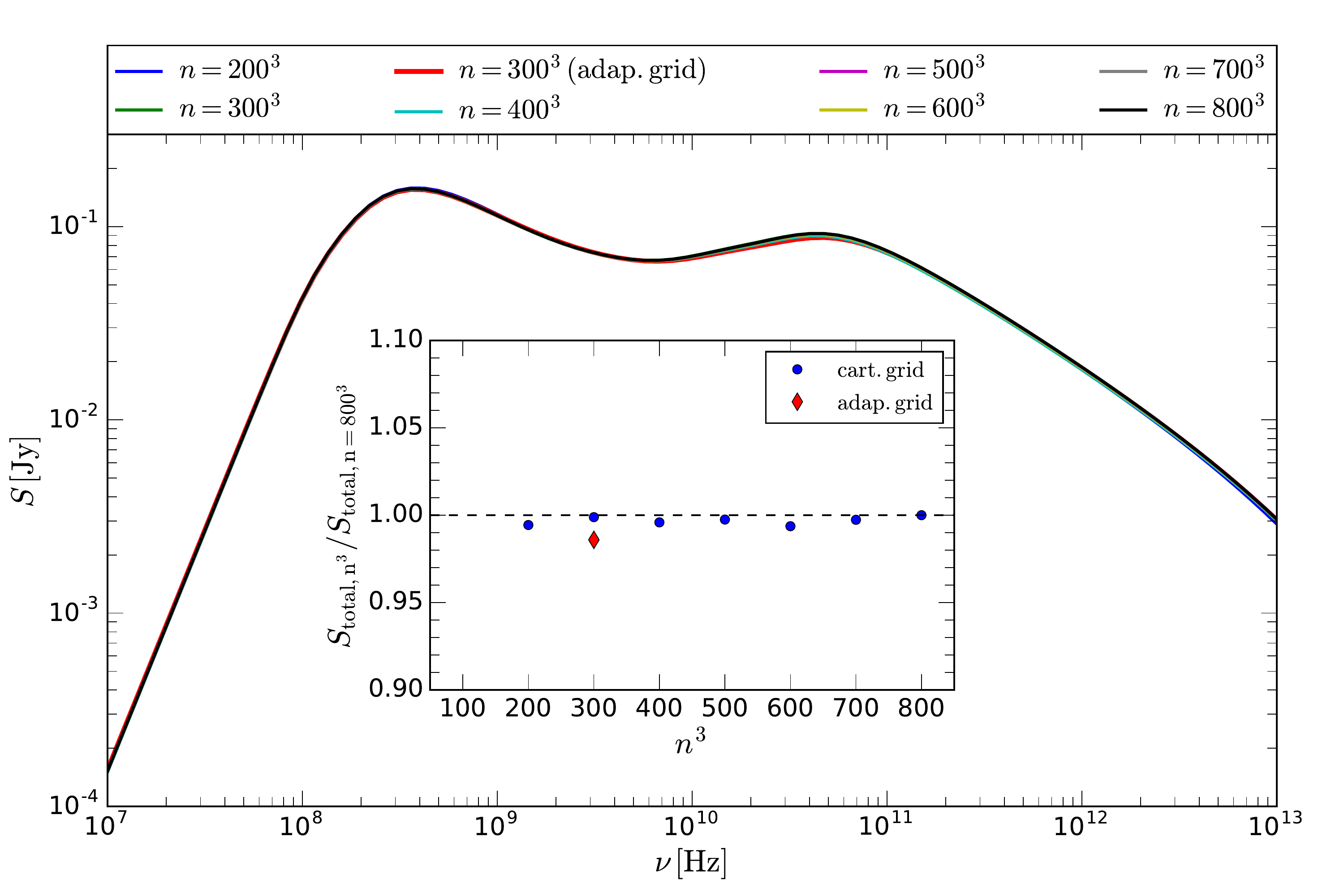}} 
\caption{Convergence test for the broadband radio spectrum for eight different resolutions}
\label{convspec} 
\end{figure}

In Fig.~\ref{convspec} we show the
single-dish spectrum between $\left(10^7 - 10^{13}\right)$\,Hz for
different resolutions and the inset panel shows the variation of the
total flux density with respect to the number of grid
cells. The calculated flux density converges to a common value with a variation of 
$S_\mathrm{total,n^3}/S_\mathrm{total,n=800^3}\leq 1\%$. The increased absorption of the 
obscuring torus leads to a decrease in the flux density between $3\times10^{8}$\,Hz and 
$5\times10^{10}$\,Hz. 
The small variation see in the inset of Fig. \ref{convspec} is caused by absorption in 
obscuring torus at high frequencies $\nu>1\times10^{10}\,\mathrm{Hz}$. The absorption in 
the torus is calculated from the temperature and the density. If we increase the numerical 
resolution variations both quantities can be better resolved by the grid. This higher resolution 
can lead to lower or higher absorption along a ray depending on its path through the torus. 
As a results, there are some small variations in the emission at high frequencies. These 
variations vanish once the ray-grid exhibits the native resolution of the SRHD simulations, 
which is $800^3$.

\begin{figure*}[t!]
\centering
\resizebox{18cm}{!}{\includegraphics{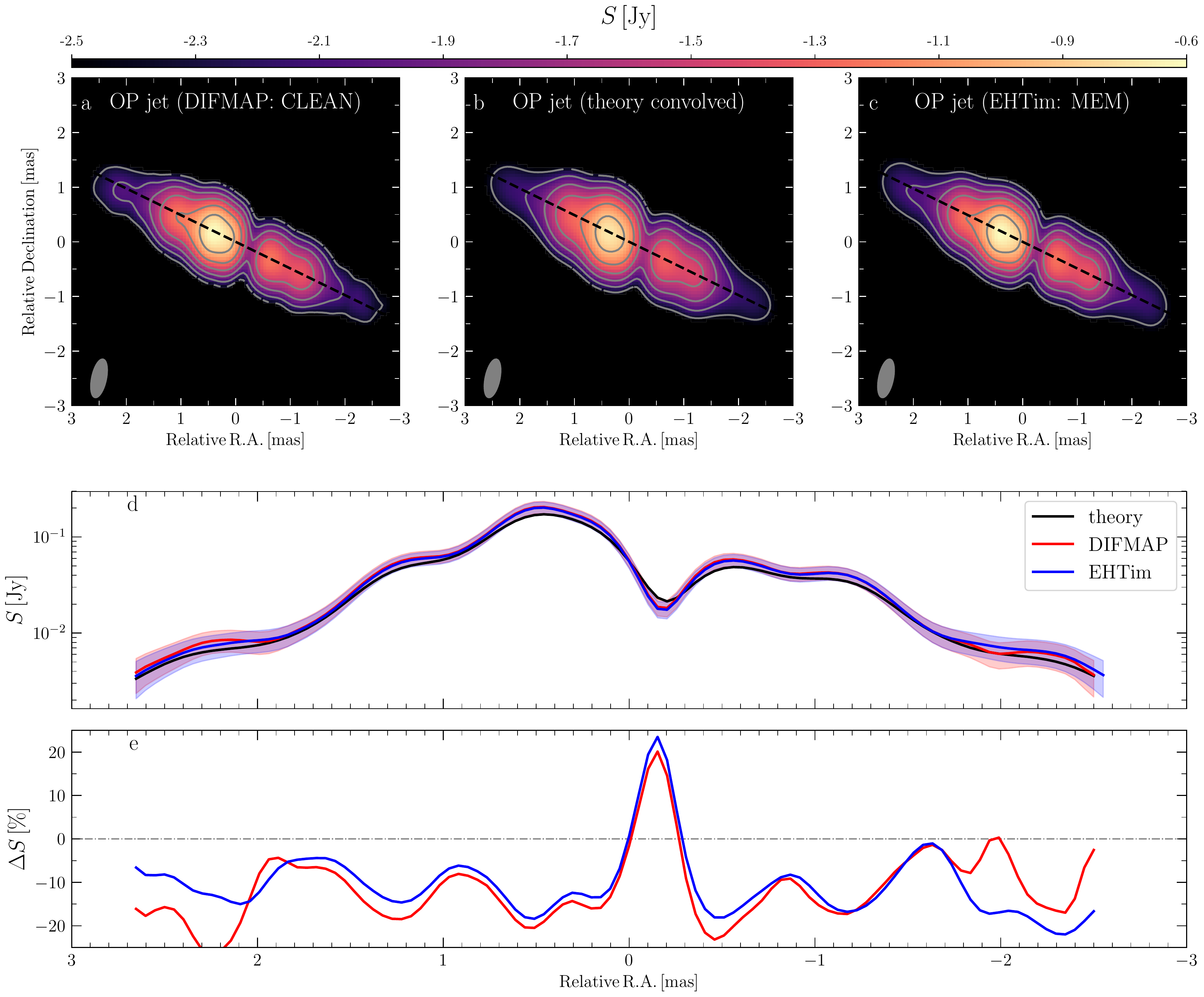}}
\caption{Result of the image reconstruction algorithm using \texttt{CLEAN} (panel a) and the \texttt{MEM} (panel c) as compared to the convolved ray-traced image (panel b). The flux density along the black dashed line in panel (a-c) is shown in panel d. The relative deviation from the theoretical flux density in precent is plotted in panel e.}
\label{beamconv} 
\end{figure*}

\subsection*{Impact of the image reconstruction and beam convolution}
In order to quantify the impact of the image reconstruction algorithm and the beam convolution on
the distribution of the flux density and therefore on the spectral index we performed a test on two
different reconstruction algorithms. In Fig. \ref{beamconv} we show reconstructed images for the OP
jet model using the \texttt{CLEAN} algorithm as implemented in \texttt{DIFMAP} (panel a) and \texttt{MEM}
algorithm from the \texttt{EHTim} package (panel c). The reconstructed images are convolved with a common 
beam of $0.27\times0.72\,\mathrm{mas}$ and flux along the jet axis is compared to the blurred infinite resolution
image (panel b). Both reconstruction algorithms provide similar results and are capable of reproducing the distribution
of the flux density along the jet axis (panel d). In panel e of Fig. \ref{beamconv} we compute flux density difference 
between the theoretical value and the reconstructed values. The average flux density difference is around 10-15\%.
This error could be added to the error budget on the flux density which will improve the $\chi^2$ values presented in 
Tables \ref{parabest} and \ref{refinedparabest}.

\subsection*{Global Millimetre VLBI Array (GMVA)}
\begin{figure}[h!]
\resizebox{\hsize}{!}{\includegraphics{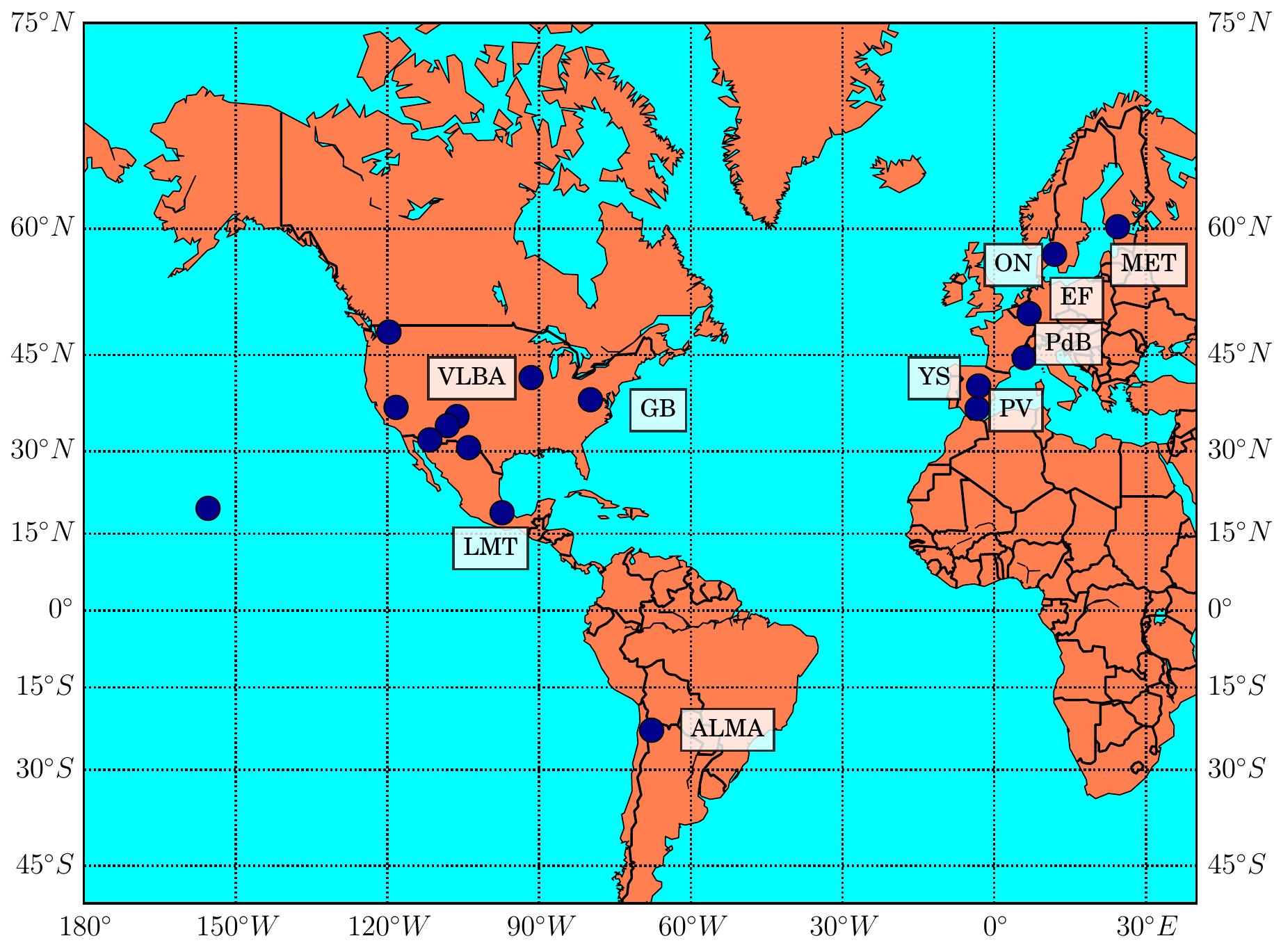}} 
\caption{Location of the different GMVA radio antennas (for station code see Table 
\ref{antennaparaGMVA})}
\label{GMVAloc} 
\end{figure}

\begin{figure}[h!]
\resizebox{\hsize}{!}{\includegraphics{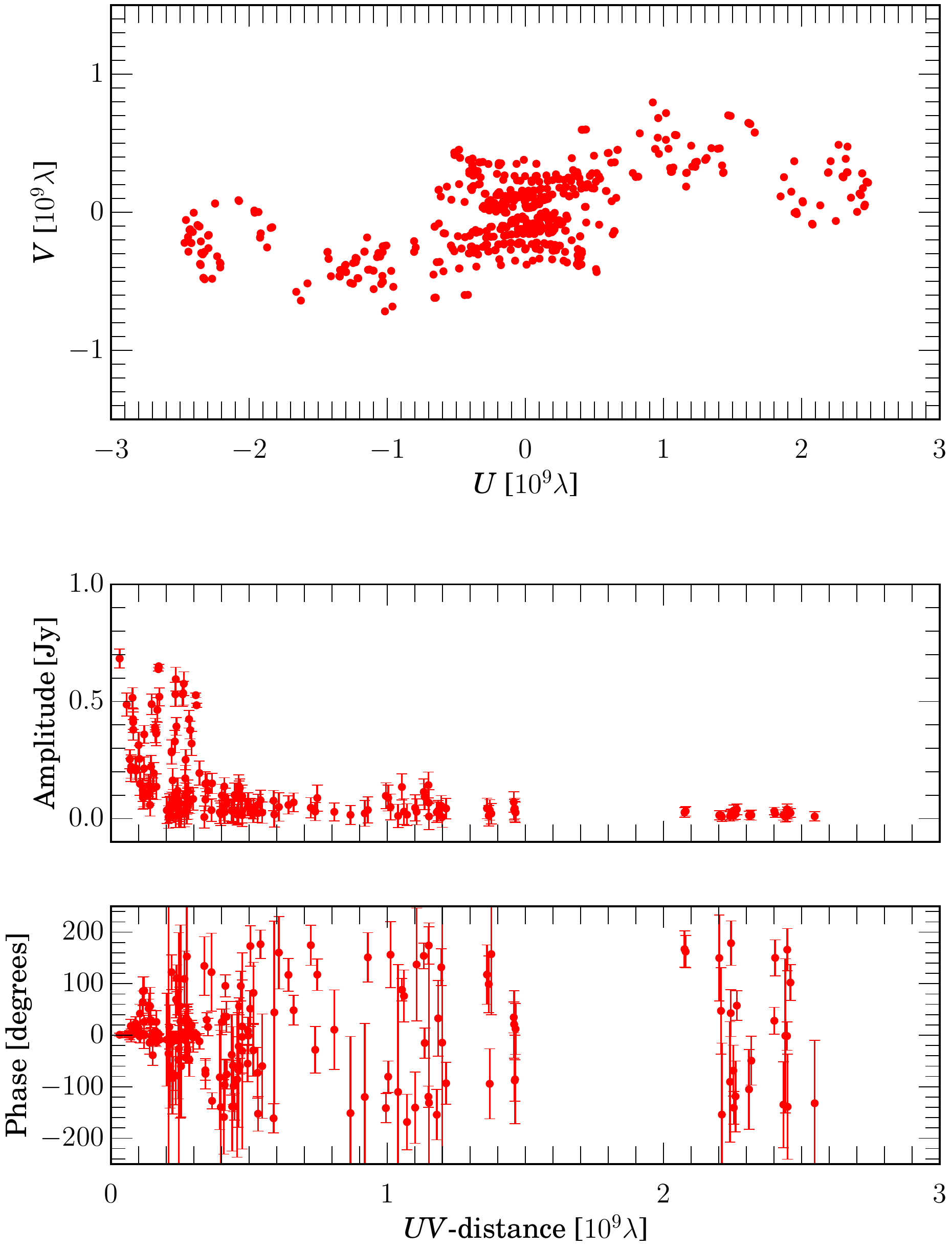}} 
\caption{Visibilities for the synthetic RadioAstron observations using the parameters listed in Table 
\ref{antennaparaGMVA} and the observing schedule described in the text. The top panel shows the 
sampling of the u-v plane, middle panel the visibility amplitude and the bottom panel the phase with uv-
distance. For reasons of clarity only every  50th data point is plotted.}
\label{visGMVA} 
\end{figure}

The GMVA\footnote{for more details see \url{https://www3.mpifr-bonn.mpg.de/div/vlbi/globalmm/}}  
operates at 86\,GHz and currently consists of the telescopes of the VLBA, the European VLBI Network 
(EVN) and ALMA. Future observations may also include the Korean VLBI stations. The system 
equivalent flux densities (SEFD) for the involved telescopes are listed in Table \ref{antennaparaGMVA} 
and the location of the telescopes can be seen in Fig. \ref{GMVAloc}. \\
For the synthetic GMVA observations of NGC\,1052 we use an integration time of 12\,s, a on-source 
scan length of 10\,min and a calibration and pointing gap of 50\,min. The observing date is set to 
October 4th 2004 and we observe NGC\,1052 from 22:00 UT until 14:00 UT on the next day. The u-v 
plane together with the amplitude and phase for the OP jet model are displayed in Fig. \ref{visGMVA}. 
\begin{table}[t!]
\caption{Used telescopes and SEFD for the GMVA observations at 86\,GHz}  
\label{antennaparaGMVA}
\centering
\begin{tabular}{l|c|c|}
\hline\hline
Name & code & SEFD [Jy]\\ 
\hline
Fort Davis & FD & 3600 \\
Los Alamos& LA & 3100 \\
Pie Town& PT & 4100 \\
Kitt Peak& KP & 4600 \\
Owns Valley& OV & 5800 \\
Brewster & BR & 3500 \\
Mauna Kea & MK & 4100 \\
North Liberty & NL & 4900 \\
Effelsberg & EF & 1000 \\
Green Bank & GB & 137 \\
Plateau de Bure & PdB & 818 \\
Pico Veleta & PV & 654 \\
Yebes & YS & 1667 \\
Onsala & ON & 5102 \\
Meths\"ahovi & MET & 17647 \\
Large Millimetre Telescope & LMT & 1714 \\
ALMA & ALMA & 68 \\
\hline
\end{tabular} 
\end{table} 
\begin{figure}[h!]
\resizebox{\hsize}{!}{\includegraphics{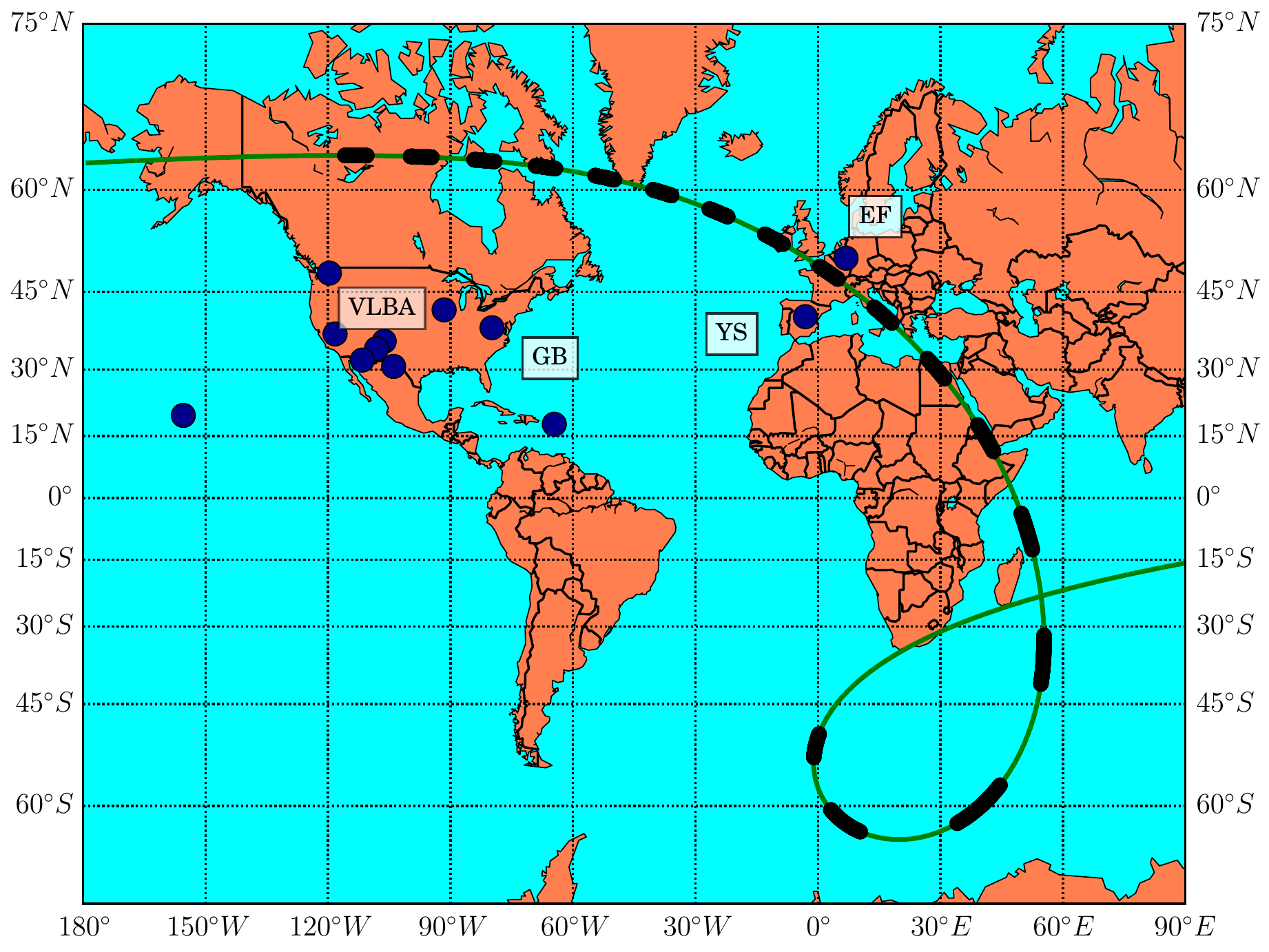}} 
\caption{Array configuration for the synthetic RadioAstron observations. The green line corresponds to 
the ground track of the RadioAstron satellite and black points indicate the individual scans. The ground 
stations are indicated by the blue points and the station codes are listed in Table \ref{antennaparaRA}.}
\label{Groundtrack} 
\end{figure}
\begin{table}[t!]
\caption{Used telescopes and SEFD for the Radioastron observations at 22\,GHz}  
\label{antennaparaRA}
\centering
\begin{tabular}{l|c|c|}
\hline\hline
Name & code & SEFD [Jy]\\ 
\hline
Fort Davis & FD & 640 \\
Hancock & HN & 640\\
Los Alamos& LA & 640 \\
Pie Town& PT & 640 \\
Kitt Peak& KP & 640 \\
Owns Valley& OV & 640 \\
Brewster & BR & 640 \\
Mauna Kea & MK & 640 \\
North Liberty & NL & 640 \\
St. Croix & SC & 640\\
Effelsberg & EF & 90 \\
Green Bank & GB & 20 \\
Yebes & YS & 200 \\
SpektR & RS & 30000 \\
\hline
\end{tabular} 
\end{table} 
\begin{figure}[h!]
\resizebox{\hsize}{!}{\includegraphics{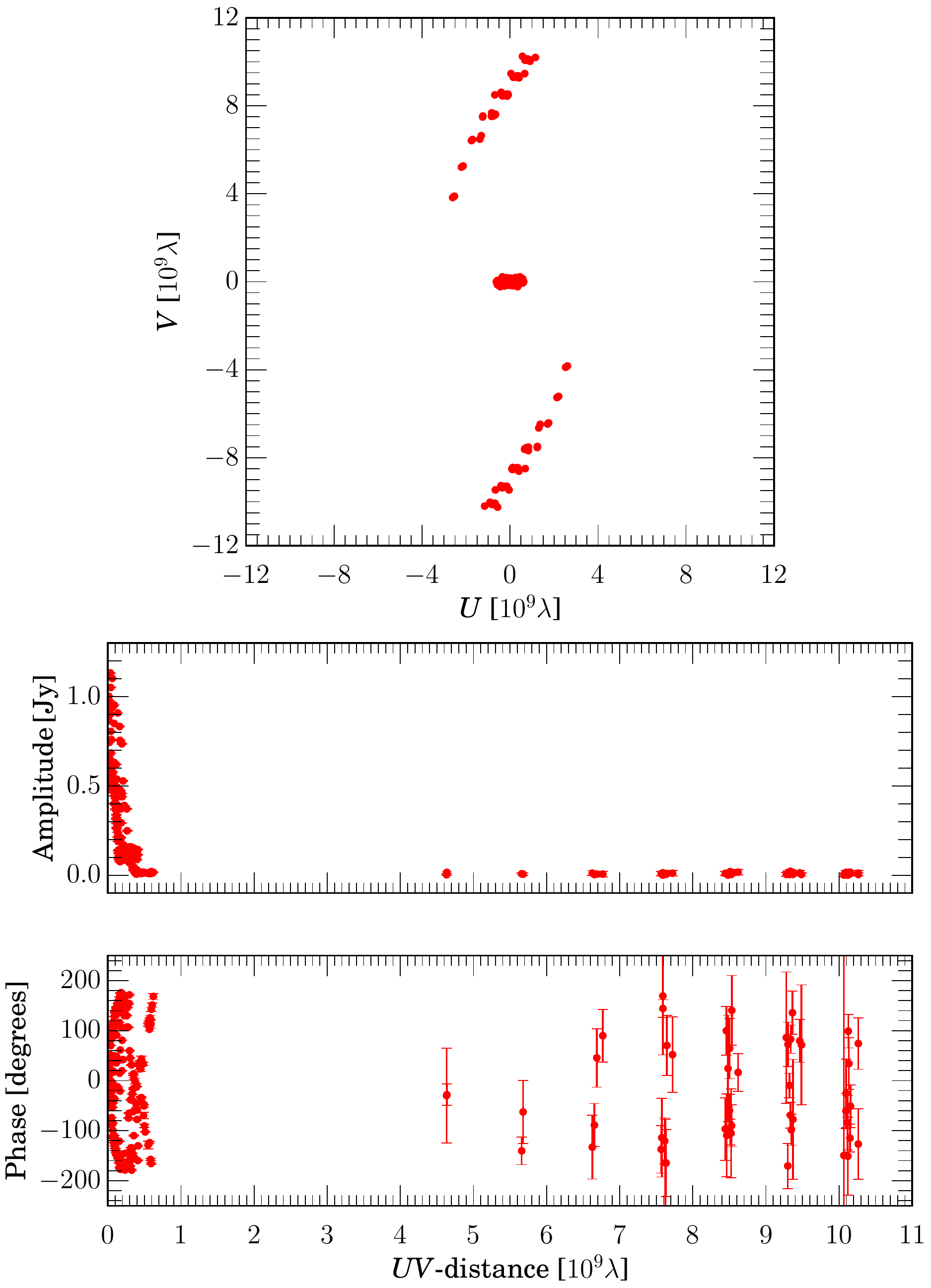}} 
\caption{Visibilities for the synthetic RadioAstron observations using the parameters listed in Table 
\ref{antennaparaRA} and the observing schedule described in the text. The top panel shows the 
sampling of the u-v plane, middle panel the visibility amplitude and the bottom panel the phase with uv-
distance. For reasons of clarity only every  50th data point is plotted.}
\label{RAvis} 
\end{figure}
\subsubsection*{Spektr-R and ground array at 22\,GHz}
The space based satellite Spektr-R extents the baselines up to 10 Earth radii and provides $\mu as$ 
resolution. The SEFD of the ground stations and the space antenna are listed in Table 
\ref{antennaparaRA}. The observing schedule for our synthetic RadioAstron observations of NGC\,1052 
is the following: 30\,min on source-scans and 95\,min off-source with an integration time of 10\,s. The 
observations are performed on the 17th of October from 17:00 UT until 18th of October 10:00 UT.
This time span corresponds to the periastron transit of the satellite and the ground track of the satellite 
together with the ground array can be seen in Fig. \ref{Groundtrack}. The solid green line is the ground 
track of RadioAstron and the black points on top of the green line corresponds to the on-source scans. 
Notice, that due the change in the orbital velocity of the satellite during the observations the spacial 
extent of the black points on top of the green lines shrinks. The location of the ground stations are 
marked by blue circles (see Table \ref{antennaparaRA} for station codes). The u-v plane and the visibility 
amplitude and phase for the OP jet model are plotted in Fig. \ref{RAvis}. Notice the large extent up to 
$12\times10^9\lambda$ in North-South direction which leads to a  beam of $160\,{\rm \mu as} \times 20\,
{\rm \mu as}$.
\subsection*{Parameter recovering tests and MCMC results}
\begin{figure*}[h!]
\centering
\resizebox{18cm}{!}{\includegraphics{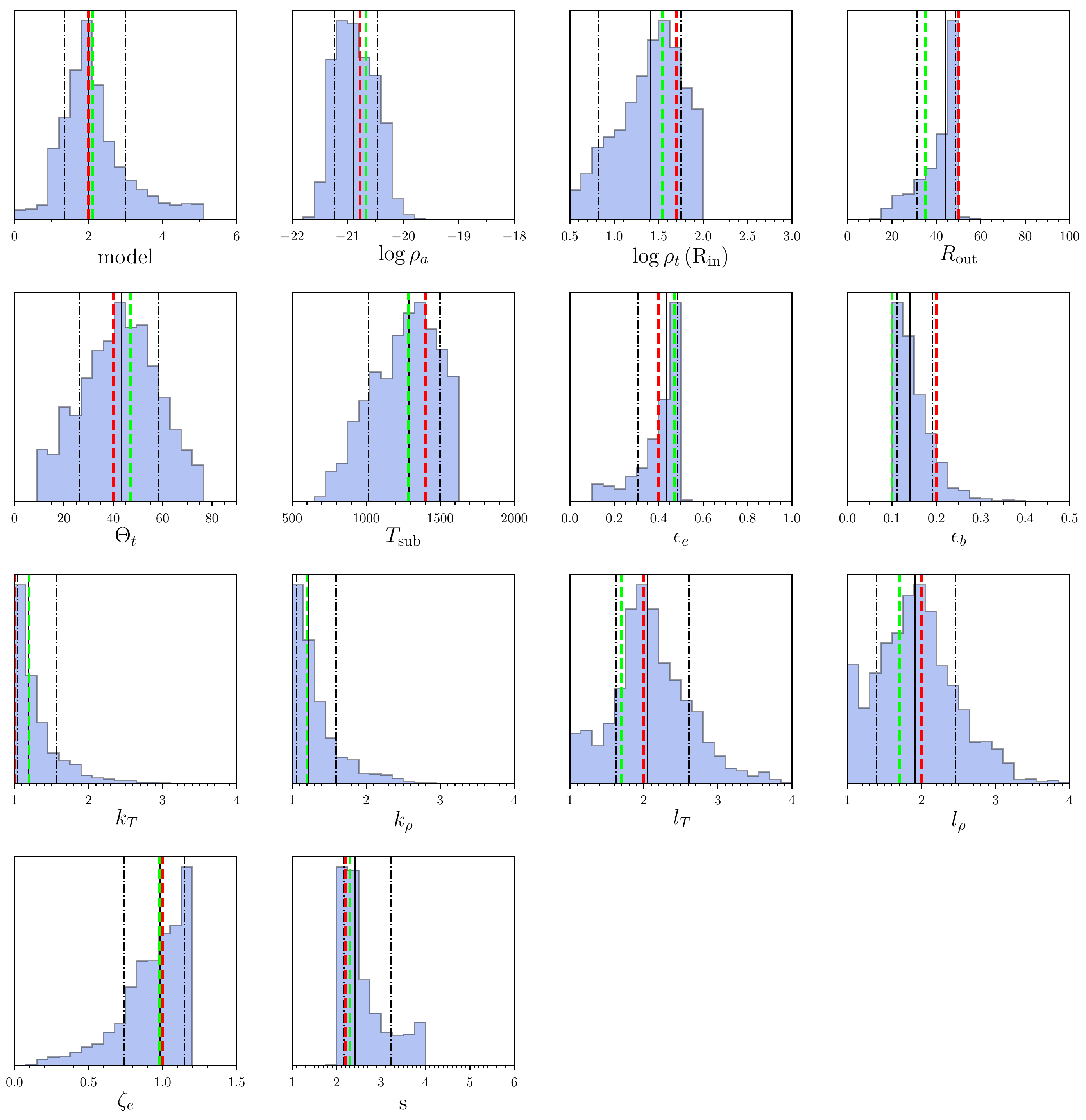}}
\caption{Results for the parameter recovering test for the MCMC simulation and the PSO. The histogram 
show the distribution of the parameters obtained from the MCMC run, solid black line indicates mean 
value and the dash-dotted line refer to $1\sigma$ 
standard deviation. The red dashed lines correspond to the parameters of the injected reference model 
and the green dashed lines to the best position recovered by the PSO.}
\label{pararecover} 
\end{figure*}

To test and explore the capabilities of your numerical optimisation we perform a parameter recovering 
test using the reference model given in Table \ref{paraemsyn} and methods described in Sect. 
\ref{opt}. We inject the reference model into our end-to-end pipeline and investigate the ability to recover 
the injected parameters. {For the PSO we 
use 400 particles, 200 outer iterations  and 5 inner iterations\footnote{Due to the constrain handling in \texttt{ALPSO} the iterations are split into inner  and outer iterations see Sect. 4 in \citet{Jansen:2011} for details.} refer and for the MCMC we employ 
400 random walkers and 1000 iterations.} These number lead to similar number of 
total iterations, namely $4\times10^{5}$. The results of the parameter recovering test can be 
seen in Fig. \ref{pararecover} and in Table \ref{pararecovertab}.
All histograms show clear maxima and injected model values are within $1\sigma$ of the 
recovered values.\\

\begin{figure*}[h!]
\centering
\resizebox{18cm}{!}{\includegraphics{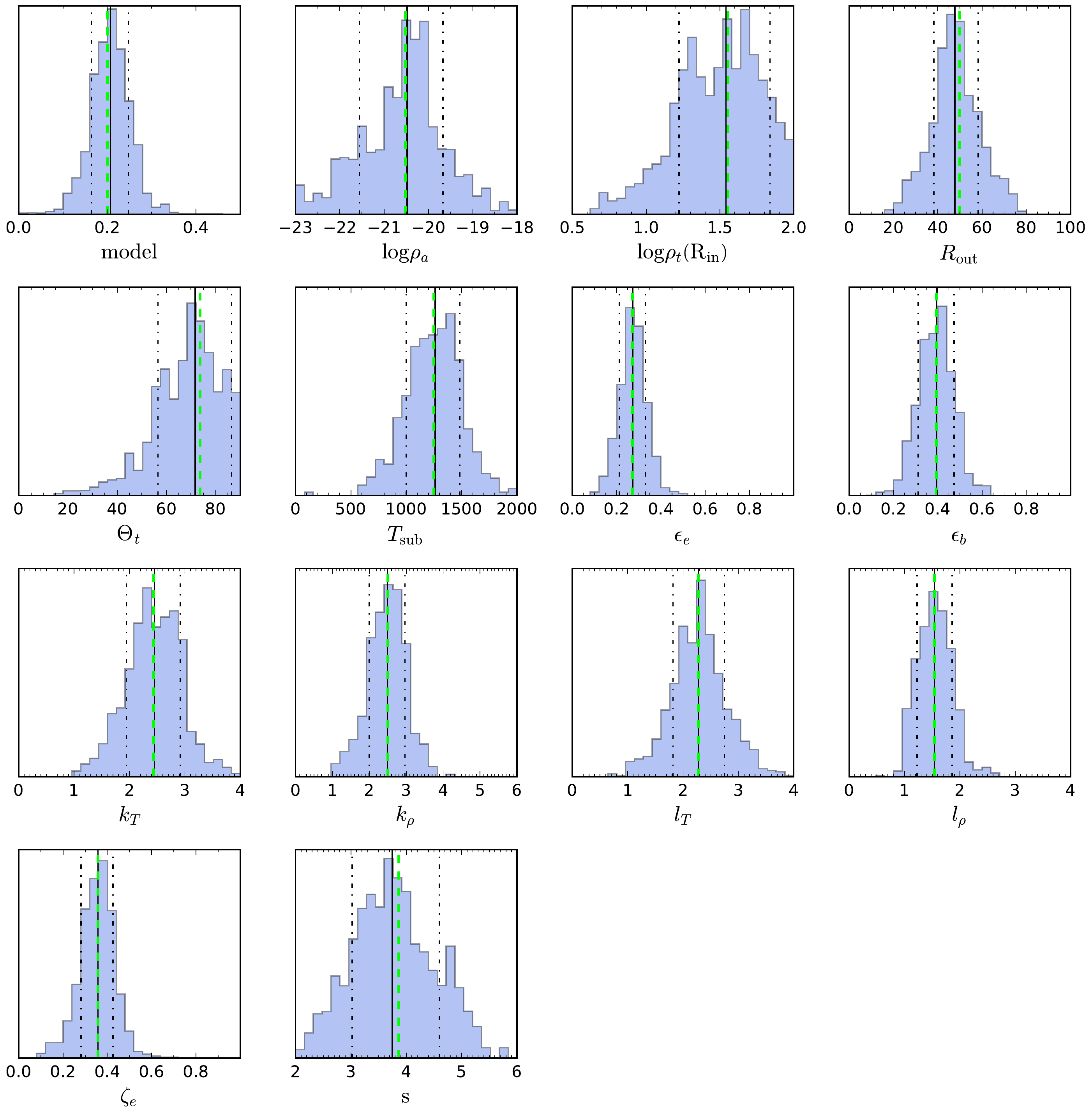}}
\caption{Results for the MCMC simulation for the OP jet model with the PSO 
best position as starting point and a standard deviation of 50\%. The histograms show the distribution of 
the parameters obtained from the MCMC run. The solid black line indicates mean value and the dash-
dotted line refer to $1\sigma$ standard deviation computed from the MCMC results. The green dashed 
lines correspond to the best position recovered by the PSO.}
\label{pararecoverOP} 
\end{figure*}

\begin{figure*}[h!]
\centering
\resizebox{18cm}{!}{\includegraphics{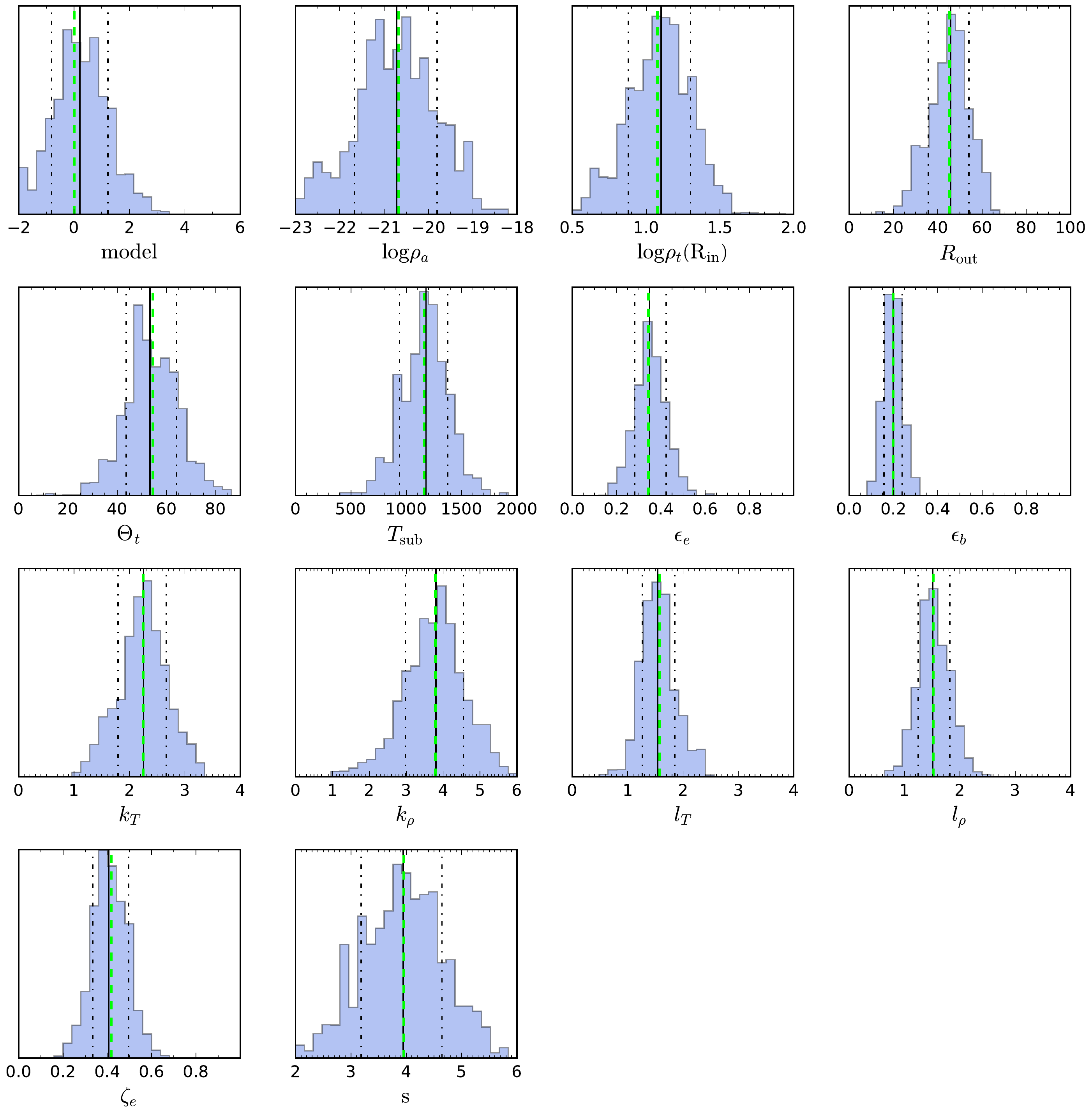}}
\caption{Results for the MCMC simulation for the PM jet model with the PSO 
best position as starting point and a standard deviation of 50\%. The histograms show the distribution of 
the parameters obtained from the MCMC run. The solid black line indicates mean value and the dash-
dotted line refer to $1\sigma$ standard deviation computed from the MCMC results. The green dashed 
lines correspond to the best position recovered by the PSO.}
\label{pararecoverPM} 
\end{figure*}

The spread around the optimal solution found for the OP and PM jet models are presented in Tables 
\ref{pararecovertabOP} and \ref{pararecovertabPM} and the histograms for the distribution can be found 
Fig. \ref{pararecoverOP} -- \ref{pararecoverPM}.
\begin{table*}[t!]
\caption{Injected parameters for the reference model and recovered parameter using PSO and MCMC}  
\label{pararecovertab}
\tiny
\centering
\begin{tabular}{|l|c|c|c|c|c|c|c|c|c|c|c|c|c|c|c|c|c|}
\hline\hline
  type & $d_k$ & $m$ & $\log\rho_a$  & $\log\rho\left(R_{\rm in}\right)$ & $R_{\rm out}$ & $\Theta$ & 
$T_{\rm sub}$  &$\epsilon_e$ & $\epsilon_B$ & $k_{\rm T}$ & $k_{\rho}$ & $l_{\rm T}$ & $l_{\rho}$ & $
\zeta_e$   & s\\ 
     & & & [g/cm$^3$] & [$\rho_a$] & $\left[R_\mathrm{j}\right]$ &  [$^\circ$] & [K]  & &  &  & &   &  & &\\ 
\hline

ref. model &2.5  &2 & -20.78 & 1.70 & 50.00 & 40.00 & 1400.00 & 0.40 & 0.20 & 1.00 & 1.00 & 2.00 & 
2.00 & 1.00 & 2.20\\
\hline \hline
PSO/GA    &2.5  &2 & -20.67 & 1.54 & 35.00 & 47.00 & 1283 & 0.47 & 0.10 & 1.20 & 1.20 & 1.70 & 1.70 
& 0.98 & 2.30\\

\hline

MCMC (mean)&2.5  &2 & -20.89 & 1.41 & 44.18 & 43.35 & 1300 & 0.44 & 0.14 & 1.19 & 1.22 & 2.05 & 
1.91 & 0.98 & 2.41 \\
MCMC ($1\sigma$) & -- & -- & $_{-0.35}^{+0.43}$ & $_{-0.59}^{+0.35}$ & $_{-12.99}^{+4.53}$ & $_{-17.03}^{+15.07}$ & $_{-276.06}^{+208.48}$ & $_{-0.13}^{+0.05}$ & $_{-0.03}^{+0.05}$ & $_{-0.15}^{+0.38}$ & $_{-0.16}^{+0.37}$ & $_{-0.42}^{+0.56}$ & $_{-0.52}^{+0.54}$ & $_{-0.24}^{+0.16}$ & $_{-0.25}^{+0.81}$\\

\hline
\end{tabular} 
\end{table*} 
\clearpage
\newpage
\begin{table*}[t!]
\caption{MCMC estimates for the OP jet model}  
\label{pararecovertabOP}
\tiny
\centering
\begin{tabular}{|l|c|c|c|c|c|c|c|c|c|c|c|c|c|c|c|c|c|}
\hline\hline
  type & $d_k$ & $m$ & $\log\rho_a$  & $\log\rho\left(R_{\rm in}\right)$ & $R_{\rm out}$ & $\Theta$ & 
$T_{\rm sub}$  &$\epsilon_e$ & $\epsilon_B$ & $k_{\rm T}$ & $k_{\rho}$ & $l_{\rm T}$ & $l_{\rho}$ & $
\zeta_e$   & s\\ 
     & & & [g/cm$^3$] & [$\rho_a$] & $\left[R_\mathrm{j}\right]$ &  [$^\circ$] & [K]  & &  &  & &   &  & &\\ 
\hline

PSO/GA & 1.5 &2 & -20.52 & 1.55 & 50 & 74 & 1249& 0.27 & 0.39 & 2.44 & 2.50 & 2.28 & 1.54 & 0.36 & 
3.86\\
\hline
MCMC (mean)&1.5 & 2 & -20.48 & 1.54 & 48 & 72 & 1260 & 0.27 & 0.40 & 2.45 & 2.49 & 2.28 & 1.54 & 
0.36 & 3.75\\
MCMC ($1\sigma$) & -- & -- & $_{-1.08}^{+0.81}$ & $_{-0.32}^{+0.30}$ & $_{-10}^{+11}$ & $_{-15}
^{+14}$ & $_{-260}^{+220}$ & $_{-0.06}^{+0.06}$ & $_{-0.08}^{+0.08}$ & $_{-0.51}^{+0.47}$ & $_{-0.49}
^{+0.48}$ & $_{-0.46}^{+0.46}$ & $_{-0.31}^{+0.32}$ & $_{-0.08}^{+0.07}$ & $_{-0.72}^{+0.85}$\\
\hline
\end{tabular} 
\end{table*} 
\clearpage
\newpage
\begin{table*}[t!]
\caption{MCMC estimates for the PM jet model}  
\label{pararecovertabPM}
\tiny
\centering
\begin{tabular}{|l|c|c|c|c|c|c|c|c|c|c|c|c|c|c|c|c|c|}
\hline\hline
  type & $d_k$ & $m$ & $\log\rho_a$  & $\log\rho\left(R_{\rm in}\right)$ & $R_{\rm out}$ & $\Theta$ & 
$T_{\rm sub}$  &$\epsilon_e$ & $\epsilon_B$ & $k_{\rm T}$ & $k_{\rho}$ & $l_{\rm T}$ & $l_{\rho}$ & $
\zeta_e$   & s\\ 
     & & & [g/cm$^3$] & [$\rho_a$] & $\left[R_\mathrm{j}\right]$ &  [$^\circ$] & [K]  & &  &  & &   &  & &\\ 
\hline

PSO/GA & 1 & 2 & -20.67 & 1.08 & 45 & 55 & 1161 & 0.34 & 0.20 & 2.25 & 3.79 & 1.58 & 1.52 & 0.42 & 
3.96\\

\hline
MCMC (mean)&1 & 2 & -20.72 & 1.10 & 46 & 53 & 1179 & 0.35 & 0.20 & 2.25 & 3.80 & 1.55 & 1.51 & 
0.41 & 3.95\\
MCMC ($1\sigma$) & -- & -- &  $_{-0.95}^{+0.96}$ & $_{-0.22}^{+0.20}$ & $_{-10}^{+9}$ & $_{-10}
^{+11}$ & $_{-240}^{+195}$ & $_{-0.07}^{+0.07}$ & $_{-0.04}^{+0.04}$ & $_{-0.46}^{+0.42}$ & $_{-0.79}
^{+0.75}$ & $_{-0.28}^{+0.30}$ & $_{-0.26}^{+0.31}$ & $_{-0.07}^{+0.09}$ & $_{-0.76}^{+0.70}$\\
\hline
\end{tabular} 
\end{table*} 
\end{document}